\documentclass[twocolumn,trackchanges]{aastex631}
\usepackage{silence} 
\usepackage{natbib}
\usepackage{amsmath}
\usepackage{float}
\usepackage{times}
\usepackage{enumitem}
\usepackage{etoolbox}
\usepackage{graphicx}
\usepackage{amsmath}
\usepackage{bm}
\usepackage{graphics}
\usepackage{url}
\usepackage{epsfig}
\usepackage{wrapfig}
\usepackage{ulem}
\usepackage{verbatim}
\usepackage{float}
\usepackage{lastpage}
\usepackage{CJK}
\usepackage[whole]{bxcjkjatype} 
\usepackage{CJKutf8}
\usepackage[caption=false]{subfig}

\def\lsim{\hbox{\rlap{\raise 0.425ex\hbox{$<$}}\lower 0.65ex\hbox{$\sim$}}}
\def\gsim{\hbox{\rlap{\raise 0.425ex\hbox{$>$}}\lower 0.65ex\hbox{$\sim$}}}

\def\arcsec{\hbox{$^{\prime\prime}$}}
\def\arcdeg{\mbox{$^\circ$}}

\newcommand\N[1]{{\color{blue} \bf #1}} %N8

\shorttitle{Spectropolarimetry of SN\,2023ixf}
\shortauthors{Vasylyev et al.}

\begin{document}

%***\alex\Are we still updating the light curve? How many months -- about 3? So, we should be able to tell whether it's Type II-P.
%%SV: I am not too comfortable quoting a light curve without presenting it here - I could of course if you think this is ok. It is spectroscopically similar to other Type IIPs. However, as Avishay pointed out, the absorption component of Halpha is a bit shallow. We don't have a completele light curve (hence why I would like to suggest we use GSP LCs for the follow up paper.)

\title{Early-time Spectropolarimetry of the Aspherical Type II Supernova SN 2023ixf} 
%\title{ Early-Time Spectropolarimetry of the Interacting Type II Supernova SN 2023ixf: Direct Evidence for Misalignment between the Ejecta and Circumstellar Material [experimental title]}% Early-Time Spectropolarimetry of the Interacting Type II Supernova SN 2023ixf: Direct Evidence for Misalignment between the Ejecta and Circumstellar Material}
\correspondingauthor{Sergiy~S.Vasylyev}
\author[0000-0002-4951-8762]{Sergiy~S.Vasylyev}
\email{sergiy\_vasylyev@berkeley.edu}
\affiliation{Department of Astronomy, University of California, Berkeley, CA 94720-3411, USA}
\affiliation{Steven Nelson Graduate Fellow}

\correspondingauthor{Yi Yang}
\author[0000-0002-6535-8500]{Yi Yang
\begin{CJK}{UTF8}{gbsn}
(杨轶)
\end{CJK}}\email{yiyangtamu@gmail.com}
\affiliation{Department of Astronomy, University of California, Berkeley, CA 94720-3411, USA}
\affiliation{Bengier-Winslow-Robertson Postdoctoral Fellow}

%\correspondingauthor{Alexei~V.~Filippenko}
\author[0000-0003-3460-0103]{Alexei~V.~Filippenko}
\affiliation{Department of Astronomy, University of California, Berkeley, CA 94720-3411, USA}
%\affiliation{Miller Senior Fellow, Miller Institute for Basic Research in Science, University of California, Berkeley, CA 94720, USA}
\author[0000-0002-1092-6806]{Kishore C. Patra}
\affiliation{Department of Astronomy, University of California, Berkeley, CA 94720-3411, USA}
\affiliation{Nagaraj-Noll-Otellini Graduate Fellow} 

\author[0000-0001-5955-2502]{Thomas~G.~Brink}
\affiliation{Department of Astronomy, University of California, Berkeley, CA 94720-3411, USA}
\affiliation{Wood Specialist in Astronomy}

\author[0000-0001-7092-9374]{Lifan Wang}
\affiliation{George P.\ and Cynthia Woods Mitchell Institute for Fundamental Physics $\&$ Astronomy, Texas A$\&$M University, 4242 TAMU, College Station, TX 77843, USA}

\author[0000-0002-7706-5668]{Ryan Chornock}
\affiliation{Department of Astronomy, University of California, Berkeley, CA 94720-3411, USA}

\author[0000-0003-4768-7586]{Raffaella Margutti}
\affiliation{Department of Astronomy, University of California, Berkeley, CA 94720-3411, USA}
\affiliation{Department of Physics, University of California, Berkeley, CA 94720, USA}

\author[0000-0002-3739-0423]{Elinor~L.~Gates}
\affiliation{University of California Observatories/Lick Observatory, Mount Hamilton, CA 95140}

\author[0000-0002-6523-9536]{Adam~J.~Burgasser}
\affiliation{Department of Astronomy \& Astrophysics, University of California at San Diego, La Jolla, CA 92093, USA;}

\author[0000-0002-1480-9041]{Preethi~R.~Karpoor}
\affiliation{Department of Astronomy \& Astrophysics, University of California San Diego, La Jolla, CA 92093, USA;}

\author[0000-0002-2249-0595]{Natalie~LeBaron}
\affiliation{Department of Astronomy, University of California, Berkeley, CA 94720-3411, USA}

\author[0000-0002-1420-1837]{Emma~Softich}
\affiliation{Department of Astronomy \& Astrophysics, University of California San Diego, La Jolla, CA 92093, USA;}

\author[0000-0002-9807-5435]{Christopher~A.~Theissen}
\affiliation{Department of Astronomy \& Astrophysics, University of California San Diego, La Jolla, CA 92093, USA;}
%\affiliation{NASA Sagan Fellow;}

\author[0009-0002-4843-2913]{Eli~Wiston}
\affiliation{Department of Astronomy, University of California, Berkeley, CA 94720-3411, USA}

\author[0000-0002-2636-6508]{WeiKang Zheng},
\affiliation{Department of Astronomy, University of California, 
Berkeley, CA 94720-3411, USA}
\affiliation{Eustace Specialist in Astronomy}

%\author{Please add name...}

%\author[0000-0002-3653-5598]{Avishay Gal-Yam}
%\affiliation{Department of Particle Physics and Astrophysics, Weizmann Institute of Science, Rehovot 76100, Israel}

%\author{...}
%PLEASE ADD YOUR NAME HERE
\begin{abstract}
We present six epochs of optical spectropolarimetry of the Type II supernova (SN) 2023ixf ranging from $\sim 2$ to 15 days after the explosion. Polarimetry was obtained with the Kast double spectrograph on the Shane 3\,m telescope at Lick Observatory, representing the earliest such observations ever captured for an SN. 
We observe a high continuum polarization $p_{\text{cont}} \approx 1$\% on days +1.4 and +2.5 before dropping to 0.5\% on day +3.5, persisting at that level up to day +14.5. Remarkably, this change coincides temporally with the disappearance of highly ionized ``flash" features. 
The decrease of the continuum polarization is accompanied by a $\sim 70^\circ$ rotation of the polarization position angle ($PA$) as seen across the continuum. The early evolution of the polarization may indicate different geometric configurations of the electron-scattering atmosphere as seen before and after the disappearance of the emission lines associated with highly-ionized species (e.g., \ion{He}{2}, \ion{C}{4}, \ion{N}{3}), which are likely produced by elevated mass loss shortly prior to the SN explosion. We interpret the rapid change of polarization and $PA$ from days +2.5 to +4.5 as the time when the SN ejecta emerge from the dense asymmetric circumstellar material (CSM). The temporal evolution of the continuum polarization and the $PA$ is consistent with an aspherical SN explosion that exhibits a distinct geometry compared to the CSM. The rapid follow-up spectropolarimetry of SN\,2023ixf during the shock ionization phase reveals an exceptionally asymmetric mass-loss process leading up to the explosion.
\end{abstract}

\keywords{supernovae: individual (SN\,2023ixf) --- techniques: spectroscopic}

%%%%%%%%%%%%%%%%%%%%
%%  Introduction  %%
%%%%%%%%%%%%%%%%%%%%

\section{Introduction}\label{sec:intro}
The core collapse of a red supergiant (RSG) with a zero-age main sequence (ZAMS) mass of at least 8\,$M_{\odot}$ is halted by the newly formed neutron star. The infalling material bounces off the neutron star, generating a shock ($v_{\rm sh} \approx 0.1c$) that eventually advances through the opaque stellar envelope. 
The shock dissipates due to radiative losses when the optical depth drops below $ \sim c/v_{\rm sh}$. When the radiation from the shock breaks out close to the surface of the star, it produces a bright X-ray/ultraviolet (UV) flash that lasts minutes to hours depending on the size of the exploding progenitor \citep{waxman_shock_2017}. 
Subsequently, the stellar envelope expands and cools, generating UV/optical emission on a longer timescale of a few days. The shock may also interact with circumstellar material (CSM) shed by the progenitor through mass-loss processes in the years leading up to the explosion. 

The high-energy radiation ionizes various species in the CSM (e.g., \ion{He}{2}, \ion{C}{3}, \ion{C}{4}, \ion{N}{3}, \ion{N}{4}) which recombine to form narrow emission lines. When discussing ``flash" features, we are referring to the emission lines formed as a result of interaction between the shock and the CSM (shock ionization). This is not to be confused with the flash-ionization features produced in the CSM by the breakout X-ray/UV flash, which recombine on a timescale of hours. Emission lines that persist beyond a few hours thus demand a long-lived source of ionizing photons that can be supplied by the interaction of the shock with dense CSM. Such emission lines that are produced as a result of ejecta-CSM interaction are a superposition of a $v \approx 100$\,km\,s$^{-1}$ P~Cygni profile and a broadened wing produced by electron scattering, which has a typical width of $\sim 1000$\,km\,s$^{-1}$ (see, e.g., \citealp{2014ApJ...797..118F, 2017NatPh..13..510Y}.)
The resulting emission is visible until the expanding ejecta sweep up the pre-existing CSM. The complete disappearance of the narrow emission lines generally within several days is observed in some explosions with RSG progenitors, thus implying that the CSM is confined to $r \leq 10^{15}$\,cm \citep{2017NatPh..13..510Y}. 

Spectropolarimetry, which measures the polarized flux at various wavelengths, is a remarkably effective tool for assessing the geometric properties of the SN ejecta and its CSM without spatially resolving the source. 
In addition, the polarization measured across prominent spectral lines can help characterize the distribution of individual elements within the ejecta. Comprehensive reviews of polarimetric studies of SNe have been provided by \citet{wang_spectropolarimetry_2008},  \citet{branch_supernova_2017}, and \citet{patat_introduction_2017}.
The fundamental principle of this method is straightforward. Polarized photons are produced via Thomson scattering in SN atmospheres, with the electric field of the scattered photon perpendicular to the plane of scattering. Any deviation from spherical symmetry would result in an incomplete cancellation of the electric-field vectors. 
Because the SN is observed as a point source and the light reaching a detector is an integration of all the electric-field vectors, any asymmetries would yield a net polarization greater than zero. Therefore, polarimetry can provide critical information about the geometric configuration shortly after the SN explosion. 

The ability to acquire and analyze the earliest spectra of core-collapse supernovae (CCSNe) has become feasible thanks to the alert stream of transients provided by contemporary wide-field sky surveys that operate on a nightly cadence, such as the Zwicky Transient Facility (ZTF; \citealp{2019PASP..131a8002B}). 
At these early stages, conventional photometry and spectroscopy offers only limited insights into the structures of the ejecta and their interaction with the CSM, projecting and smearing the geometry into a single dimension of radial velocity. 
%\textcolor{red}{YY: `traditional...' > `conventional photometry and spectroscopy offer only limited clue on the structures of the ejecta and its interaction with any circumstellar matter, projected and smeared into the single dimension of radial velocity.'}%SV2YY: I'm trying to think of better wording here. Will come back to later.

The central portion of the emission profiles of the highly-ionized ``flash" features is primarily shaped by recombination photons from the CSM. As a result, it typically exhibits a width $\sim 100$\,km\,s$^{-1}$, reflective of the velocity of the CSM. Consequently, the narrow core of the feature remains essentially unpolarized as the emission lines from recombination form typically above the electron-scattering photosphere. 
%OLD: Polarimetry, which measures the deviation from spherical symmetry, provides critical information about the geometric configuration starting soon after the SN explosion.

On the other hand, the wings of the ``flash" emission features, which originate from electron scattering within the highly ionized CSM and are represented by a symmetric Lorentzian profile \citep{chugai_broad_2001,chugai_type_2004,2018MNRAS.475.1261H}, can produce polarization if the CSM distribution deviates from spherical symmetry.
Consequently, the degree of polarization ($p$\%) and the $PA$ across the electron-scattering wings can help map out the geometric distributions of the associated elements in the CSM. A comparison of the polarization spectrum over the continuum and such ``flash" features also serves to evaluate whether the axially symmetric explosion of the progenitor and its pre-explosion mass-loss process align along a similar symmetry axis. 

%\textcolor{red}{YY: Please move this paragraph above the previous one.}
%Spectropolarimetry, which measures the polarized flux at varied wavelengths, is a remarkably effective tool for assessing the geometric properties of the SN ejecta without spatially resolving the source. In addition, the polarization measured across prominent spectral lines can help characterize the distribution of individual elements within the ejecta. Comprehensive reviews of polarimetric studies of SNe have been provided by \citet{wang_spectropolarimetry_2008, branch_supernova_2017, patat_introduction_2017}.
%The fundamental principle of this method is straightforward. Polarized photons are produced via Thomson scattering in SN atmospheres, with the electric field of the scattered photon perpendicular to the plane of scattering. Any deviation from spherical symmetry would result in an incomplete cancellation of the electric-field vectors. As the SN is observed as a point source and the light reaching a detector is an integration of all the electric-field vectors, any asymmetries would yield a net polarization greater than zero. 

Over the past few decades, polarimetric measurements have suggested that the core collapse of massive stars is generally aspherical. 
However, SNe with an optically thick H-rich envelope were observed to have relatively low polarization ($p \leq 0.2$ \%) during the phase of hydrogen recombination, suggesting the presence of an almost spherical H-rich envelope \citep{wang_spectropolarimetry_2008}. 
The light curve of the SN remains on a ``plateau" that typically lasts a few months after the explosion, as the hydrogen recombination becomes the dominant emission source. 

For example, time-series polarimetry of the Type IIP SN\,2004dj showed negligible continuum polarization ($p \leq 0.1$\%) during its plateau phase, followed by a rapid increase to $p = 0.56$\% when
the outer H-rich envelope recombined and the inner aspherical helium core was revealed \citep{leonard_non-spherical_2006}. 
This physical picture of an aspherical explosion surrounded by a spherical hydrogen envelope has become ubiquitous among many other SNe\,II (see, e.g., \citealp{2001PASP..113..920L, chornock_large_2010, nagao_aspherical_2019, dessart_polarization_2021}). 

However, recent studies have also identified a $\sim 1$\% continuum polarization intrinsic to the Type IIP/IIL SN\,2013ej \citep{2017ApJ...834..118M, 2021MNRAS.505.3664N} and Type IIP SN\,2021yja \citep{2023arXiv230306497V} from as early as $\sim 2$--3 weeks after the explosion. 
The polarization spectra of SN\,2021yja follow a dominant axis as displayed on the Stokes $q-u$ plane, indicative of a well-defined axial symmetry starting from the outer layer of its hydrogen envelope and in contrast to the previous low early continuum polarization of H-rich SNe. 
Such a high degree of large-scale asphericity during the early phase of the explosion is likely induced by an interaction process preserving a well-defined symmetry axis, potentially involving a binary companion or a significant concentration of CSM in the equatorial plane \citep{2023arXiv230306497V}.

SN\,2023ixf was discovered on 2023-05-19 17:27 (UTC dates are used throughout this paper) / MJD 60083.73 in the spiral galaxy Messier\,101 at a clear-band magnitude of 14.9 \citep{2023TNSTR1158....1I}. A prediscovery clear-band detection about 11.4\,hr earlier at $16.0 \pm 0.3$\,mag was reported by \citet{filippenko_filippenko_2023}.
The earliest detection of the SN was MJD $60082.83 \pm 0.02$ \citep{mao_onset_2023} at mag 17.7. The time of the first light is estimated to be MJD 60082.75  \citep{2023arXiv230606097H}, which will be used throughout this paper.
A classification spectrum obtained at 2023-05-19 22:23 / MJD 60083.93 contains a series of emission lines of H, He, C, and N \citep{2023TNSAN.119....1P}, centering at zero velocity relative to the reported redshift $z=0.000804$ of its host galaxy \citep{2022ApJ...934L...7R}. These prominent emission features, superimposed on a blue continuum, are characteristic of SNe\,II observed within mere hours to a few days following the explosion \citep{2023TNSAN.119....1P}. Several works have constrained the progenitor to be a RSG, with evidence of periodic pre-explosion variability \citep{kilpatrick_sn2023ixf_2023,jencson_luminous_2023,soraisam_sn_2023}. %RSG progenitor 
Throughout our analysis, all spectra have been corrected to the rest frame.

This paper is organized as follows. In Section~\ref{sec:obs}, we summarize our spectropolarimetric observations of SN\,2023ixf and discuss systematic errors associated with instrumentation and contamination from interstellar polarization. 
Section~\ref{sec:results} describes the temporal evolution of the polarization degree and polarization position angle. A qualitative discussion of the major spectropolarimetric properties of SN\,2023ixf and an interpreted geometry of the SN ejecta and the CSM is presented in Section~\ref{sec:discussion}. We summarize our results in Section~\ref{sec:conclusion}. 

\section{Spectropolarimetric Observations and Data Reduction}~\label{sec:obs}
SN\,2023ixf was observed using the Kast double spectrograph on the Shane 3\,m reflector at Lick Observatory \citep{miller_ccd_1988, miller_stone_1994}. 
A dense spectropolarimetric sequence was obtained, consisting of six epochs of observations that spanned from +1.4 to +14.5 days after the explosion. To our knowledge, the first epoch is the earliest spectropolarimetry of any known SN to date. 

Observations and data reduction were carried out following the description provided by \citet{patra_spectropolarimetry_2021}. Telluric lines were removed through comparison with the flux spectrum of the standard-star BD\,+262606 \citep{oke_secondary_1983}. 
The 300\,line\,mm$^{-1}$ grating, a GG455 order-blocking filter, and the 3$\arcsec$-wide slit were adopted, resulting in a spectral resolving power of $R \approx 380$, corresponding to the size of a resolution element of 18\,\AA\ at a central wavelength of 6800\,\AA.
A log of spectropolarimetric observations is presented in Table~\ref{tbl:specpol_log}. 

%YY20230701\begin{table*}
%\begin{center}
%\caption{Journal of spectropolarimetric observations of SN\,2023ixf.} 
%    \begin{tabular}{cccccccc}
%	\hline 
%	\hline
%	UT Date & MJD$^b$ & Phase$^a$ & Airmass & Avg. Seeing &  Exposure Time$^c$   \\ 
%	(MM-DD-YYYY)&   & (days) & & (arcsec) & (s)   \\ 
%	\hline 
%    05-20-2023 & 60084.18 & 1.4 & 1.12--1.44 & 1.2 & $600 \times 4 \times 4$ \\
%    05-21-2023 &60085.23& 2.5 & 1.06& 1.2 & $600 \times 4 \times 3$ \\
%    05-22-2023 &60086.24& 3.5 & 1.05 & 1.2 &  $600 \times 4 \times 1$\\
%    05-23-2023 &60087.35& 4.6 & 1.15 & 1.2 &  $600 \times 4 \times 1$ \\
%    05-28-2023 &60092.20& 9.5 & 1.05 & 1.2 &  $600 \times 4 \times 4$ \\
%   06-02-2023 &60097.26& 14.5 & 1.06 & 1.2 &  $200 \times 4 \times 2$\\
%	\hline 
%\end{tabular}\\
%%\flushleft
%{$^a$}{Days after the estimated time of first light on MJD~60082.75 (UTC 18 May 2023).} \\
%{$^b$}{MJD is given as the start time of the CCD exposure.}\\
%{$^c$}{Exposure time of a single exposure $\times$ 4 retarder-plate angles $\times$ number of loops. Wavelength range for Kast is 4550--9800\,\AA.} 
%\label{tbl:specpol_log}
%\end{center}
%\end{table*} 

In order to characterize the instrumental polarization, nightly observations of the low-polarization standard star HD\,212311 were carried out with the same polarimetry optics used to observe SN\,2023ixf. 
Following the expression adopted by \citet{2022MNRAS.509.4058P}, the average of the intensity-normalized Stokes parameters derived for the unpolarized standard are minimal ($q=Q/I<0.05$\%, $u=U/I<0.05$\%), 
thus confirming the negligible instrumental polarization and high stability of the Kast spectropolarimeter. The typical systematic uncertainties expected from the instrument and the reduction procedure are $\sigma_{q}, \sigma_{u} \approx 0.1$\% \citep{leonard_is_2001}.  
We carried out a sanity check by comparing repeated observations of SN\,2023ixf within a single night. Our findings suggest that these systematic uncertainties are accurately characterized by $\sigma_{q} \approx 0.1\%$ and $\sigma_{u} \approx 0.1\%$, within the continuum wavelength range of 5600--6400\,\AA.
These systematic errors are well above the statistical uncertainties calculated for the stokes parameters. 
A detailed discussion of the systematic effects expected from the Kast spectropolarimeter can be found in the Appendix of \citet{leonard_is_2001}. 
%\textcolor{red}{YY: You commented out my two suggestions above, both I believe to be valid. Could you please reconsider? SV2YY: I'm applying your suggestions}

\section{Results of Spectropolarimetry Observations}~\label{sec:results}
\begin{figure*}[!htb]
   \begin{minipage}[t]{0.5\textwidth}
     \centering
     \includegraphics[trim={0.5cm 2.5cm 0.5cm 2.5cm},clip,width=1.03\linewidth]{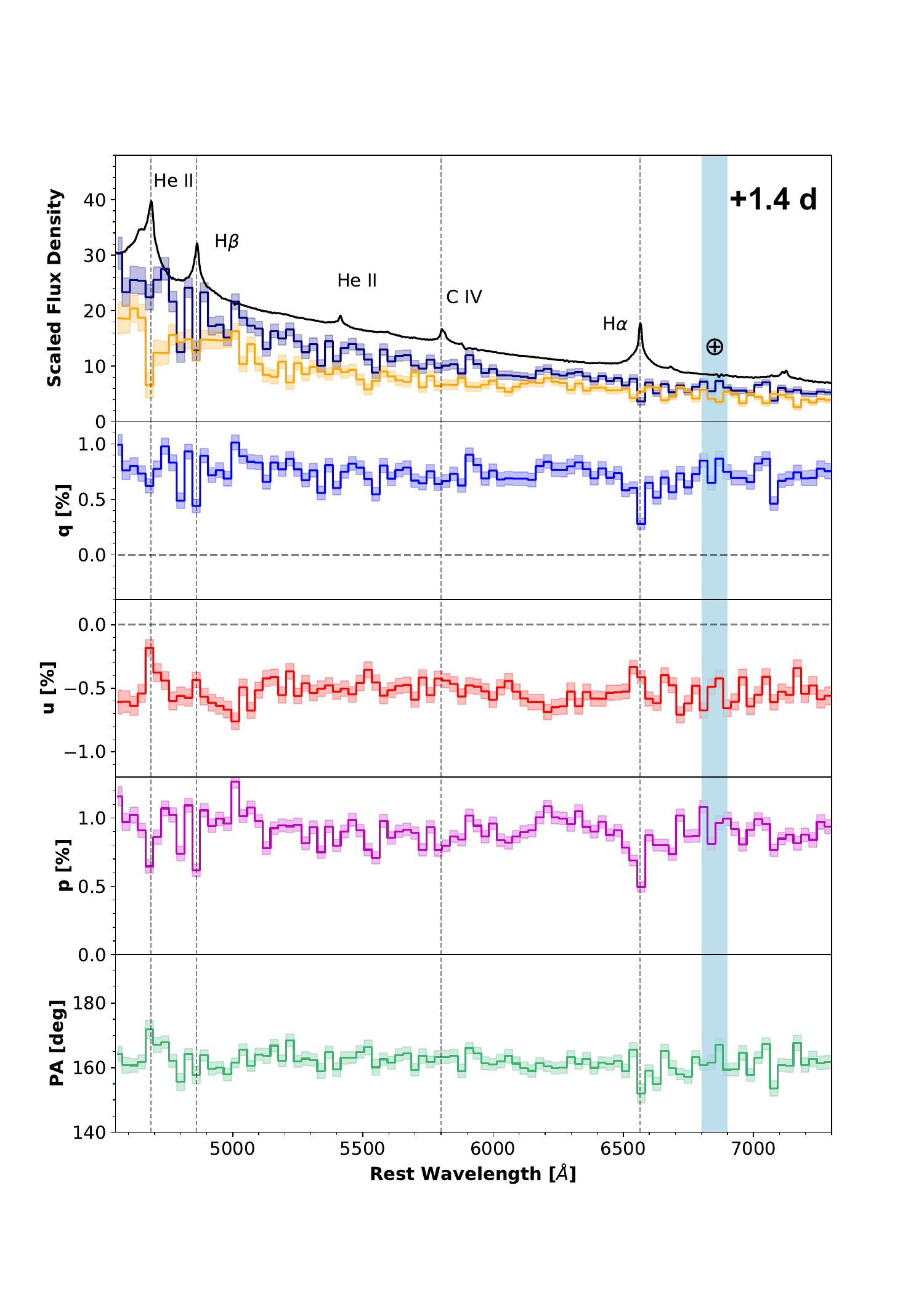}
     \caption{
     Spectropolarimetry of SN\,2023ixf on day +1.4 (epoch 1) relative to an estimated time of first light on MJD~60082.83. The five panels (from top to bottom) present
     (a) the arbitrarily scaled flux density, $f_{\lambda}$ (black line), and the polarized flux densities as indicated by the quantities $|q| \times f_{\lambda}$ (navy histogram) and $|u| \times f_{\lambda}$ (orange histogram). Major spectral features at zero velocity are labeled and marked by vertical grey lines; (b,c) the fractional Stokes parameters $q$ and $u$, respectively; (d) the degree of polarization ($p$); and (e) the polarization position angle ($PA$). In panels (b)–-(e), the data have been rebinned to 30\,\AA\ for clarity. No ISP correction has been applied. Light-blue-shaded vertical bands mark the regions of major telluric absorptions (which have been almost entirely removed through comparisons with standard-star spectra). Light shaded histogram in each panel indicates the 1$\sigma$ uncertainties.} %contamination.
     ~\label{fig:pol_ep1}
   \end{minipage}\hfill
   \begin{minipage}[t]{0.5\textwidth}
     \centering
     \includegraphics[trim={0.5cm 2.5cm 0.5cm 2.5cm},clip,width=1.03\linewidth]{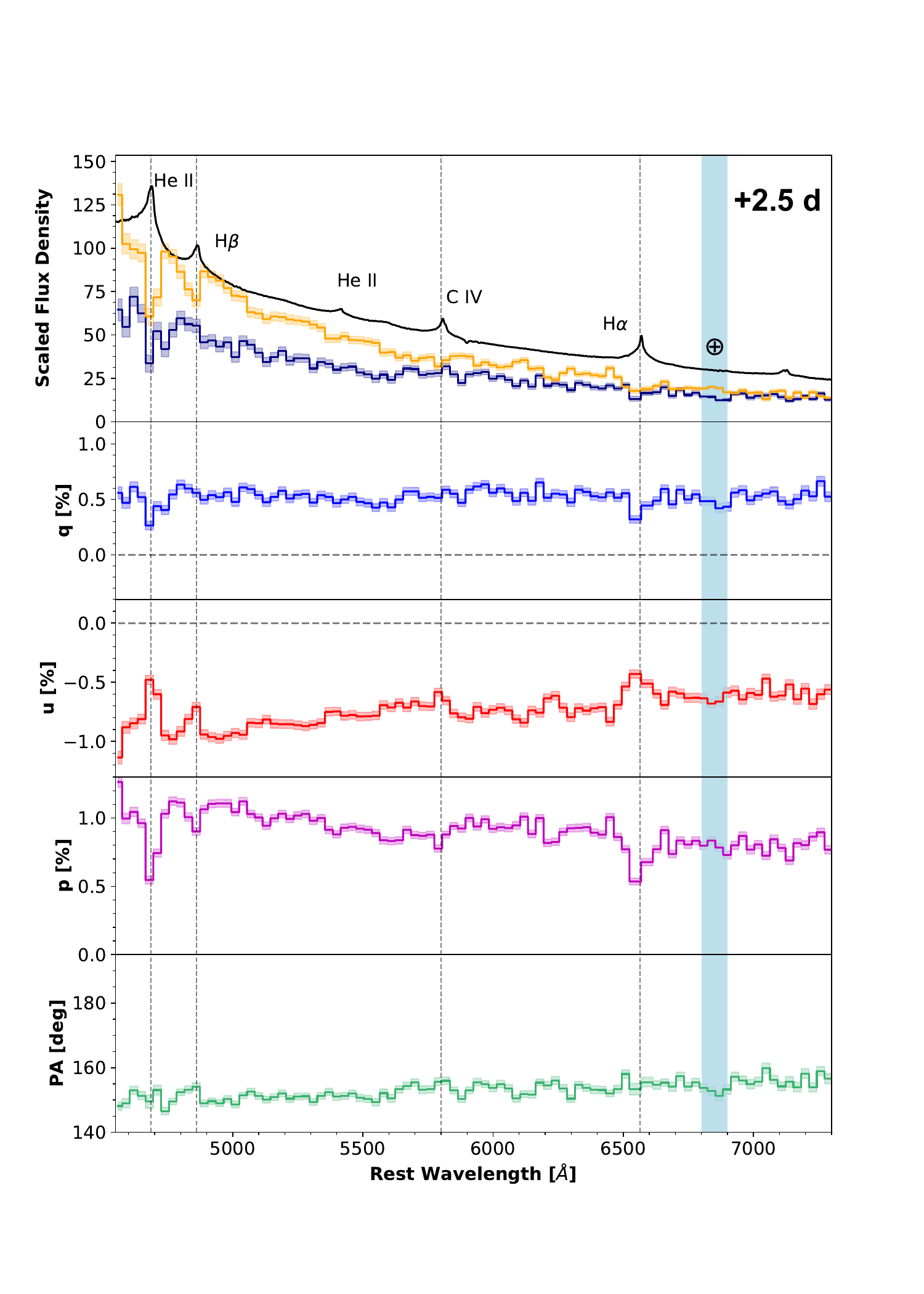}
     \caption{Same as Figure~\ref{fig:pol_ep1}, but for day $+$2.5 (epoch 2).}~\label{fig:pol_ep2}
   \end{minipage}
\end{figure*}

The flux-density spectrum of SN\,2023ixf at day +1.4 was derived from an average of two sets of Stokes $I$ parameters, acquired from two spectropolarimetric observations at MJD~60084.20 and 60084.39. The polarization measurements for both epochs were found to be consistent within the systematic errors and thus were combined in our discussion. As indicated in Figure~\ref{fig:pol_ep1}, the spectrum showcases a sequence of narrow emission lines superimposed on a blue continuum. These include the Balmer lines, \ion{He}{2} $\lambda\lambda$4686, 5411; \ion{C}{4} $\lambda\lambda$5801, 5812; \ion{N}{3} $\lambda\lambda$4634, 4641, 4687; \ion{N}{4} $\lambda\lambda$4068, 7115; and \ion{He}{1} $\lambda\lambda$5876, 6678. The spectral signatures are comparable to those in the earliest classification spectrum obtained 6\,hr earlier at MJD~60083.93 \citep{2023TNSAN.119....1P}. 
%\textcolor{red}{
%The flux density spectrum of SN\,2023ixf at day +1.4 was obtained through an average of two sets of the Stokes $I$ parameters acquired through two sets of spectropolarimetric observations at MJD 60084.xx and 60084.yy. As labeled in Figure~\ref{fig:pol_ep1}, the spectrum displays a series of narrow emission lines, including the Balmer lines, \ion{He}{2}$\lambda$4686, 5411, 5876, 6683; \ion{C}{5}$\lambda$5801, 5812; \ion{N}{3}$\lambda$4634, 4641, 4687; \ion{N}{4}$\lambda$4068, 7115; \ion{He}{1}$\lambda$6678, 7065, superimposed on a blue continuum. The spectral signatures are similar to the earliest classification spectrum that was acquired xx hours earlier at MJD 60083.93 \citep{2023TNSAN.119....1P}.
%}

Additionally, the total-flux spectrum obtained from the second set of spectropolarimetry at MJD~60084.39 also shows a significantly decreased strength in the \ion{N}{3} $\lambda\lambda$4634, 4641 profile, along with an apparent weakening of the Balmer lines. A more detailed discussion of the spectral evolution on an hourly timescale within the first two days after the SN explosion, as the SN ejecta sweep through the pre-existing CSM, will be presented in a separate paper (Zimmerman et al., in prep.).

\subsection{Interstellar Polarization}~\label{sec:isp}
The intrinsic polarization of the SN\,2023ixf explosion could be contaminated by the polarization induced by interstellar aspherical dust grains along the SN-Earth line of sight. This interstellar polarization (ISP) is produced by the dichroic extinction caused by paramagnetic dust grains that have been partially aligned by a large-scale magnetic field (see \citealp{1951ApJ...114..206D}).  
The total host-galaxy reddening, $E(B-V)_{\text{tot}}$, to SN\,2023ixf is expected to be low given the high Galactic latitude ($b \approx 59.8^{\circ}$) and favorable location of the SN in its face-on host.
The Galactic and the host-galaxy reddening of SN\,2023ixf can be estimated as $E(B-V)_{\rm MW}=0.0074$\,mag based on the extinction map given by \citet{2011ApJ...737..103S}, and $E(B-V)_{\rm Host}=0.031$\,mag as derived from the equivalent width of the \ion{Na}{1}\,D absorption lines at the redshift of the host galaxy \citep{2023arXiv230607964S}, respectively. 
Adopting an $R_V=3.1$ extinction law \citep{1989ApJ...345..245C}, we set an empirical upper limit for the ISP caused by dust along the SN-Earth line of sight, $p_{\rm ISP} \lesssim9\% \times E(B-V)_{\rm total} = 0.35\%$ 
\citep{1975ApJ...196..261S}. The maximum possible ISP, $p_{\rm ISP,max} = 0.35$\%, is substantially smaller than the observed polarization at all times. Therefore, in the subsequent discussion, we do not include any ISP correction. %Add sentence referencing later discussion

%\textcolor{blue}{RC: I think this ends too abruptly. It could use a few sentences to state that this maximum ISP value is much smaller than the observed polarization (and cannot explain its time evolution), point to figure 2 (with my suggested ISP circle), and then a statement that you will ignore ISP in the following discussion and will revisit that at the end.}

\subsection{Temporal Evolution of the Continuum Polarization}~\label{sec:cont_pol}
\begin{figure*}
    \centering
    \includegraphics[width=0.65\textwidth]{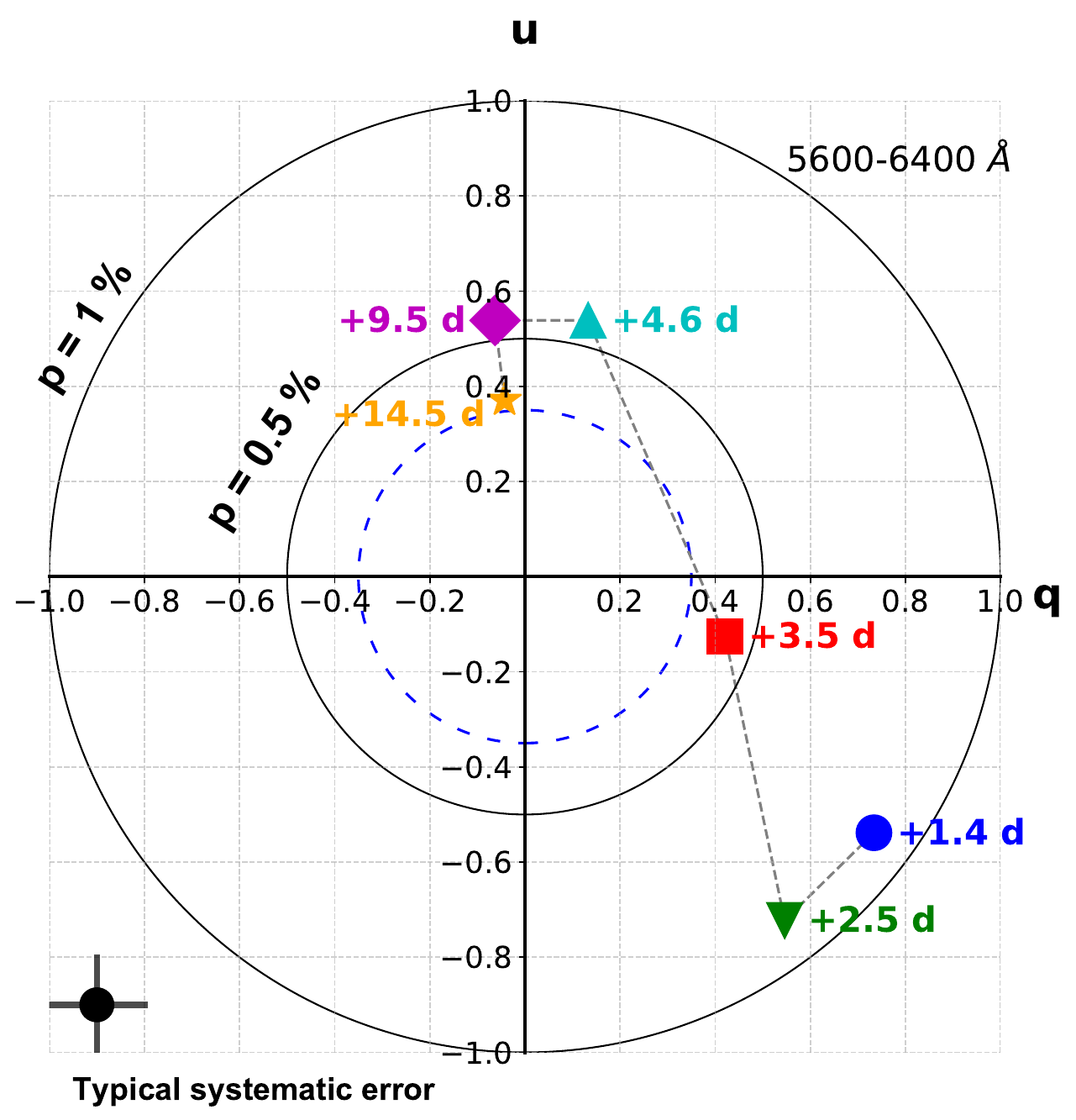}
    \caption{Temporal evolution of the continuum polarization of SN\,2023ixf from days +2 to +15 after the explosion displayed in the Stokes $q-u$ plane. The continuum is calculated over the wavelength range 5600--6400\,\AA. The phases of the measurements are relative to an estimated time of first light of MJD~60082.75 \citep{2023arXiv230606097H,2023arXiv230607964S}. The scale of the lower-left cross represents the typical systematic error as estimated from adopting various parameters during the extraction of the individual spectrum obtained during the polarization measurement. The outer and the inner solid circles mark the 1.0\% and the 0.5\% contours of the polarization level, respectively. The dashed blue circle indicates the upper limit on the ISP, $p_{\rm ISP} \lesssim 0.35\%$.  
    }
    \label{fig:story}
\end{figure*}

In Figure~\ref{fig:story}, we illustrate the evolution of the continuum polarization of SN\,2023ixf from days +1.4 to +14.5 in the Stokes $q-u$ plane. For each epoch, we compute the error-weighted mean Stokes parameters within the wavelength range 5600--6400\,\AA, a region free of prominent spectral lines. In order to facilitate easier comparison with any future work, we confirm that the continuum polarization measured across 6900--7200\,\AA\ agrees with the aforementioned range within the systematic uncertainties. The continuum polarization is also measured across the wavelength range 6700--8400 \AA\ as a consistency check. The continuum polarization measured across both 5600-6400 and 6700--8400 \AA\ also agree within the systematic uncertainties.
%The continuum polarization is also measured across the wavelength range 6700--8400 \AA\ as a consistency check and is tabulated in Table~\ref{tbl:contpol}. 

\begin{table}
\caption{Continuum polarization of SN\,2023ixf$^a$} 
\begin{tabular}{lcccc}
	\hline 
	\hline
	 Phase$^b$ & $q_{\texttt{cont}}$ & $u_{\texttt{cont}}$ & $p_{\texttt{cont}}$ &$PA_{\texttt{cont}}$$^c$   \\ 
	 (days) & &  & (\%) & (deg)   \\ 
	\hline 
%    09-11-2021 && & Kast & 1.9-2.1 & 3 & 3 & 360 \\
    1.4 & 0.73 & $-0.54$ & 0.91 & 162(3) \\
    2.5 & 0.55 & $-0.72$ & 0.91 & 153(3)  \\
    3.5 & 0.42 & $-0.12$ & 0.43 & 172(7)\\
    4.6 & 0.13 & 0.54& 0.56 & 218(5)  \\
    9.5 & $-0.06$ & 0.54 & 0.54 &  228(5) \\
    14.5 & $-0.04$ & 0.37& 0.38 &  228(8)\\
    
	\hline 
\end{tabular}\\
%\flushleft
{$^a$}The continuum polarization is calculated over the range 5600--6400\,\AA. Systematic errors for $q$ and $u$ ($\sim 0.1$\%) are well above the statistical uncertainties. \\
{$^b$}{Days after the estimated time of first light on MJD~60082.75 (18 May 2023).} \\
%{$^b$}{MJD is given as the start time of the CCD exposure.}
{$^c$}{Systematic uncertainties for $PA$ are presented in parentheses.}~\label{tbl:contpol}
\end{table}

As shown in Figure~\ref{fig:story} and Table~\ref{tbl:contpol}, the continuum polarization observed on both days +1.4 and +2.5 is roughly consistent, at $p \approx 1$\% and at a similar $PA \approx 160^{\circ}$. 
These estimates are in good agreement with the $BVRI$ polarimetric observations of SN\,2023ixf with the Liverpool Telescope on MJD~60085.07 \citep{Maund_2023_pol}. 
Subsequently, a significant and rapid change in polarization was observed from days +2.5 to +4.6. 
At later epochs, the polarization of the SN settled to $p \approx 0.5$\% and $PA \approx 225^{\circ}$, where the polarized signal has been predominantly observed along the $u$ direction in the Stokes $q-u$ plane as shown in Figure~\ref{fig:story}. 
The rapid change of the polarization during $\sim 3$--5 days after the SN explosion implies a significant alteration in the geometric configuration of the electron-scattering atmosphere at particular phases. 
The rather constant continuum polarization measured after day $\sim 5$ suggests that the geometry of the SN has settled to a distinct configuration compared to the earlier phases. 

%\textcolor{red}{
%As shown in Figure~\ref{fig:story} and Table~\ref{tbl:contpol}, the continuum polarization observed at both days +2 and +3 are roughly agreed at $p \sim$1\% and at a similar $PA \sim 165^{\circ}$. After that, a significant and rapid change in polarization was observed from days +3 to +5. At later epochs, the polarization of the SN has settled to p$ \sim$ 0.5\% and PA$ \sim 225^{\circ}$, in which the polarized signal has been seen mostly along the $u$ direction in the Stokes $q-u$ plane as shown in Figure~\ref{fig:story}. The rapid evolution of polarization within the first $\sim$5 days after the SN explosion implies a significant alteration in the geometric configuration of the electron-scattering atmosphere at particular phases. The rather constant continuum polarization measured after day $\sim$5 suggests that the geometry of the SN has settled to a distinct configuration compared to the earlier phases.
%}

\begin{figure*}
    \centering
\includegraphics[trim={1.0cm 3.0cm 1.0cm 3.0cm},clip,width=0.95\textwidth]{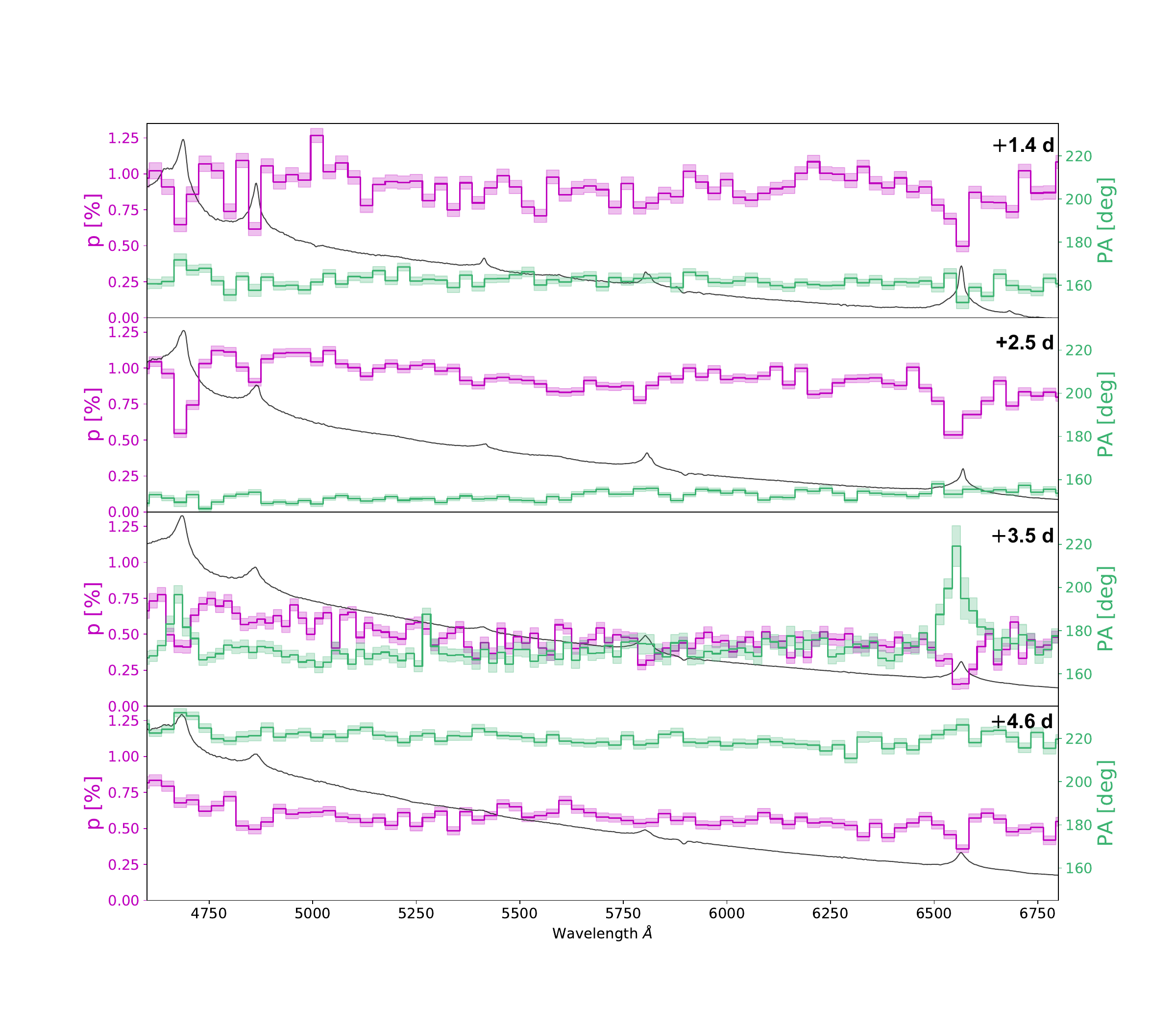}
    \caption{Temporal evolution of the polarization spectrum of SN\,2023ixf from days +1.4 to +4.6 as shown from the top to bottom panels. In each panel, $p$ and $PA$ are shown by the magenta and green histograms, respectively. 
The gray line presents the arbitrarily-scaled flux-density spectrum. 
%Major flash features such as H$\alpha$, H$\beta$, and \ion{He}{2}+\ion{N}{3} are marked by vertical lines at zero velocity. 
At days +1.4 and +2.5, the ``flash" features (FWHM $\approx 1000$\,km\,s$^{-1}$) display $PA \approx 160^{\circ}$ that is consistent with the continuum. 
At day +3.5, the continuum polarization has decreased from $\sim 1.0$\% to $\sim 0.5$\%. \ion{He}{2} $\lambda$4686 and the Balmer lines are broadened (FWHM $\approx 1000$\,km\,s$^{-1}$). The $PA$ over the emission lines also exhibits a progressive modulation toward $\sim 220^{\circ}$ towards the inner wings. 
After the CSM that produces the ``flash" features has been swept away by the expanding ejecta at day +4.6, the continuum polarization has settled to a level of 0.5\% and $PA \approx 220^{\circ}$ after a further rotation in the Stokes $q-u$ plane compared to day +3.5 (see Figure~\ref{fig:story}). 
The rapidly evolving polarization across the continuum, the Balmer lines, and \ion{He}{2} $\lambda$4686 during the disappearance of the ``flash" features implies that aspherical ejecta break out through an aspherical CSM within the first five days after the SN explosion. 
}~\label{fig:ipth}
\end{figure*}
%%\begin{figure*}
%%    \centering
%%    \includegraphics[width=0.5\textwidth]{figures/xxx.png}
%%    \caption{
%%YY: Perhaps adding a figure based on the following caption: 
%%Stokes parameters of SN\,2023ixf displayed on the $q-u$ plane as measured from the wavelength range of 4500--8900 \AA. The data have been rebinned to 20\AA\ for the epochs displayed and labeled. The wavelength of each bin can be interpreted from the color bar at the top. In each panel, the large symbol displays the continuum polarization deduced from the wavelength range 5600--6400 \AA\ as shown in Figure~\ref{fig:story}. The blue and color circles represent the measurement within $\pm$4,000 km s$^{-1}$ of the zero velocities of \ion{He}{2}$\lambda$4686 and H$\alpha$, respectively. 
%%    }~\label{fig:qu}
%%\end{figure*}

\subsection{Polarization Evolution across Emission Lines}~\label{sec:line_pol}
On both days +1.4 and +2.5, the $PA$ values across the scattering wings [with full width at half-maximum intensity (FWHM) $\approx$ 1000\,km\,s$^{-1}$] of the major flash features such as H$\alpha$ and the \ion{He}{2}+\ion{N}{3} complex are overall consistent with that measured across the continuum (see Figures~\ref{fig:pol_ep1}, \ref{fig:ipth}, and \ref{fig:pol_ep2}). Such rather flat patterns in $PA$ across the flash lines corroborate a similar geometric configuration shared by the regions that emitted the earliest continuum and created the scattering wings of the ``flash" features --- both are from an aspherical distribution of the CSM.

%\textcolor{blue}{SV: Co-Authors, Please give some feedback on this section (in \textcolor{red}{red}). We would like some feedback on this interpretation.}
An alternative method for investigating the polarization of spectral features is to compare the flux density ($f_{\lambda}$) with its polarized components, $|q|\times f_{\lambda}$ and $|u|\times f_{\lambda}$. 
The broadened wings arise from electron scattering above the recombination front of the associated ``flash" features. Thus, the polarized flux density shows the source that participates in the scattering process. In other words, its spectral shape mirrors the flux-density spectrum of the obscured source \citep{Hough_2006}. 
Therefore, when measuring the polarized flux density across the highly ionized lines, any deviation from their adjacent continuum would quantify the polarization of such features.
%\textcolor{purple}{SV2YY: In that case, what can you get from this as opposed to looking at q and u , while also plotting the flux density. It seems redundant to me but I guess it is easier to interpret. Also, we should cite Tran+ 2017 using this method as well. By the way, in that paper, it is also clear that plotting q, u and qxflam and uxflam is redundant unless you are trying to look for redshift or blueshift in the polarized flux. }

As shown in the top panel of Figure~\ref{fig:pol_ep1}, we compare the spectral profiles of the polarized flux densities $|q| \times f_{\lambda}$ and $|u|\times f_{\lambda}$ with the total flux density. We identify a deviation in polarized flux density across \ion{He}{2} $\lambda$4686 + \ion{N}{3} $\lambda\lambda$4634, 4641 from the adjacent continuum, which appears to be stronger in $|u|\times f_{\lambda}$ compared to $|q|\times f_{\lambda}$. However, no such deviation can be seen across H$\alpha$. This discrepancy between different ``flash" features can be explained by a higher electron scattering optical depth experienced by the \ion{He}{2} $\lambda$4686 + \ion{N}{3} $\lambda\lambda$4634, 4641 photons compared to the H$\alpha$ photons, which is caused by the different depth of their recombination fronts and last scattering regions --- the recombination front for the former line complex resides deeper in the CSM (\citet{groh_early-time_2014}; see Figure 4 of \citep{shivvers_early_2015}).
At day +2.5, despite the continuum polarization remaining unchanged compared to day +1.4, more significant deviations in $|q|\times f_{\lambda}$ and $|u|\times f_{\lambda}$ from the continuum can be clearly seen across both \ion{He}{2} $\lambda$4686 and H$\alpha$ (Figure~\ref{fig:pol_ep2}). The rapidly fading \ion{N}{3} $\lambda\lambda$4634, 4641 has already disappeared. We caution that interpretations of these features are speculative, as the polarization flux is not as reliable at wavelengths blueward of 5500\,\AA.  This is largely due to the decreasing detector sensitivity at shorter wavelengths. We also find that the systematic uncertainties below 5500\,\AA\ are higher ($p \approx 0.2$\%) than for the region 5500--6700\,\AA\ ($p \approx 0.1$\%).

The trend of increasing polarized line flux density continues on day +3.5. The widths of the \ion{He}{2} $\lambda$4686 feature and the Balmer lines have increased to FWHM $\approx 2000$\,km\,s$^{-1}$. The $PA$ across the wings of these prominent emission profiles has evolved into a modulation that changes progressively from $\sim 165^{\circ}$ to $\sim 225^{\circ}$ as measured toward the inner part of the wings. 
The latter $PA$ is also consistent with the continuum polarization of the SN after day +4.6, when the CSM has been swept away by the expanding ejecta. 
As the narrow emission lines are diminishing after day +4.6, the $PA$ across major spectral profiles has leveled out with the continuum.

\section{Discussion}~\label{sec:discussion}
\begin{figure*}
    \centering
    \includegraphics[width=1.0\textwidth]{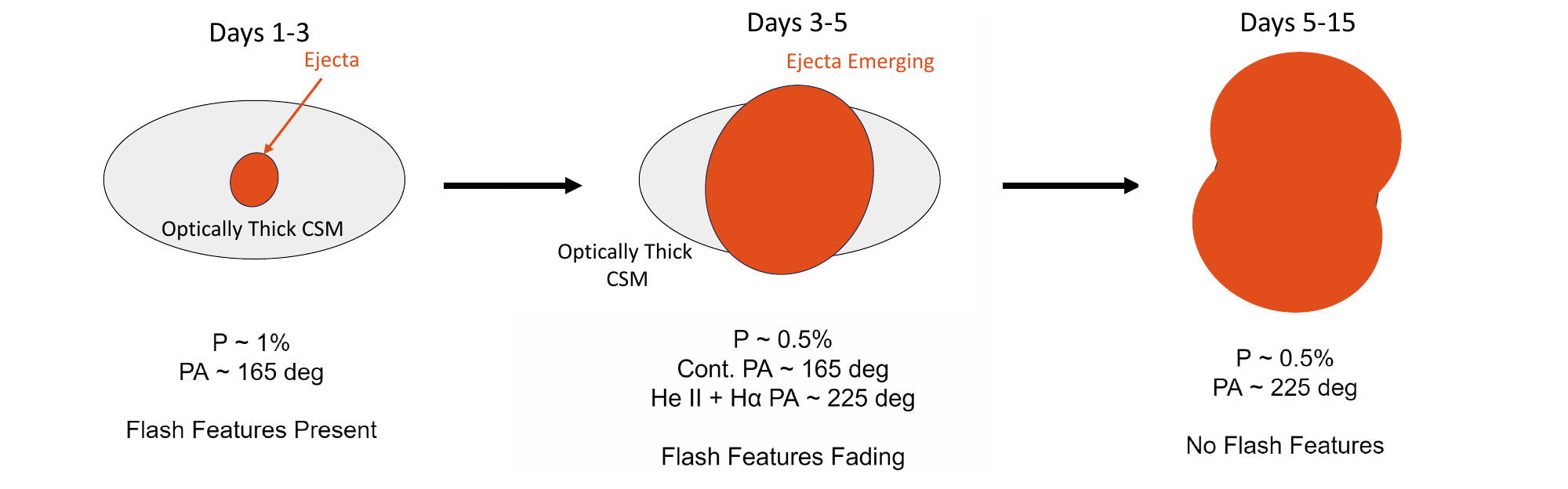}
    \caption{A simplified schematic sketch of the temporal evolution of the ejecta and CSM geometry of SN\,2023ixf as inferred from spectropolarimetry, viewed from a direction perpendicular to the plane of the sky. 
Arbitrary scales and the axis ratios of the ejecta and the CSM are used for illustration purposes. The actual axis ratio that effectively characterizes the degree of asphericity of the CSM depends on its radial density profile, electron-scattering optical depth, and viewing angle. 
{\it Left:} configuration $\lesssim +2.5$ days, when the SN ejecta are still enclosed within the outer bound of the dense, aspherical CSM. The photosphere is in a breakout region within the CSM that is enriched by pre-explosion mass loss from the progenitor; the continuum and the short-lived, narrow, highly ionized features are  from the CSM.
{\it Middle:} the transitional phase between days +2.5 and +4.6 when the aspherical SN ejecta emerge from the CSM.
{\it Right:} the expanding ejecta engulf the CSM, as indicated by the disappearance of the shock-ionized ``flash" features. 
The SN ejecta take over the CSM as the main optical emission source after day +4.6. 
%\textcolor{purple}{SV: Just need to change He I to He II.}
    }
    \label{fig:schem}
\end{figure*}

To summarize, we have observed, for the first time, the temporal evolution of the spectropolarimetric properties of a young Type II SN before its CSM has been swept away by the ejecta. 
A schematic picture to illustrate this early ejecta-CSM interaction based on the spectropolarimetry of SN\,2023ixf is sketched in Figure~\ref{fig:schem}. %The evolution of the flux spectra is presented in Figure \ref{fig:fluxes}. 

The rapid evolution of the ``flash" features during the first $\sim 3$ days after the SN explosion suggests a confined radial distribution caused by the CSM enrichment from the progenitor within a period of  
$$\Delta t_{\rm wind} \approx 100\,{\rm d}\,\frac{\Delta t_{\rm flash}}{\rm 1\,d} \frac{\rm 100\,km\,s^{-1}}{v_{\rm wind}} \frac{v_{\rm ej}}{\rm 10,000\,km\,s^{-1}},$$
\noindent
where $\Delta t_{\rm flash}$ gives the duration of the flash features, and $v_{\rm wind}$ and $v_{\rm ej}$ respectively denote the velocity of the wind and the expanding SN ejecta. Such dense CSM around SN\,2023ixf has been suggested to be confined to a radius of $\sim 5\times10^{14}$\,cm \citep{bostroem_early_2023} or $\sim 10^{15}$\,cm \citep{2023arXiv230604721J}. Furthermore, given the lack of a blueshift in the narrow absorption component of the flash features' P~Cygni profiles, the CSM is expected to be asymmetric and predominantly away from our line of sight to the SN \citep{2023arXiv230607964S}. Additionally, X-ray and radio signatures were observed for SN\,2023ixf by \citet{grefenstette_nustar_2023} and \citet{berger_millimeter_2023}, respectively, suggestive of a truncated CSM. 

The $\sim 1$\% continuum polarization observed before day +2.5 is compatible with the presence of an optically thick, asymmetric CSM. 
Within the regime of an optically thick CSM ($\tau_{\rm CSM}\approx c / v_{\rm shock} > 1$), where $v_{\rm shock}$ gives the shock velocity, the photosphere (optical depth $\tau \approx 2/3$) would reside significantly above the layer of $\tau_{\rm CSM}$. Therefore, a $\sim 1$\% continuum polarization before day +2.5 traces the geometry of the outer regions of the dense CSM while the SN ejecta are contained within it. 
%\textcolor{red}{YY: `the photons emitted by the SN ejecta are mostly stalled in'. SV: is it actually stalled or is it just that the photons take a while to diffuse or are reprocessed? The ejecta shockwave certainly isn't stalled, so I may need to think about the wording here. YY: Changed to `photons', please check now.}\textcolor{purple}{SV2YY: Why do photons diffuse out on a dynamical timescale? shouldn't this be a diffusion timescale? YY: `upon they diffuse out the ejecta on dynamic timescales'}
Considering the CSM will remain opaque to the inner ejecta, despite a decreasing $\tau_{\rm CSM}$ within the first few days, the 
polarization will still stay close to an asymptotic limit when the optical depth is larger than $\tau_{\rm CSM} \approx 3$ (see, e.g., Figure 1 of \citealt{hoflich_asphericity_1991}). 

The synchronous disappearance of the flash features and the drift of the continuum polarization in the Stokes $q-u$ plane as shown in Figure~\ref{fig:story} suggest a large-scale transformation of the geometry as the CSM is swept away by the SN ejecta. 
In addition to the continuum polarization at day +3.5, which has been measured to be located midway between days +3 and +5 in the Stokes $q-u$ plane, conspicuous evolution of the polarization spectrum across the broadened Balmer lines and \ion{He}{2} $\lambda$4686 can be identified (see Figure~\ref{fig:ipth}). 
This broadening of the emission profiles is also seen in the  high-resolution spectra obtained at similar phases \citep{2023arXiv230607964S}. The $PA$ displays a steady increase from $\sim 165^{\circ}$ to $\sim 225^{\circ}$ from the electron-scattering wings to the line rest wavelength of H$\alpha$. Interestingly, the former is consistent with the continuum $PA$ at day +3.5, while the latter seems to agree with the continuum $PA$ as measured after day +4.6. 

A possible explanation for the observed rotation in $PA$, seen solely across the H$\alpha$ and \ion{He}{2} wings at day +3.5 and coinciding with the continuum $PA$ after day +4.6, is that the H and He lines originate from the outer regions of the elongated SN ejecta emerging from the CSM. As the shock-driven ejecta break through the CSM, a significant shift is observed in both the continuum and the scattering wings, settling to a roughly constant continuum polarization and $PA$ after day +4.6.  

%Systematic errors induced by various choices made during the data reduction, for instance, variations in the wavelength solution below 5500 \AA, owing to the lack of lamp spectra and the selection of aperture sizes, have been evaluated by repeating the reduction across a parameter space grid. Additionally, consecutive observations of polarization standards on the Kast spectrograph have shown the systematic error floor to be $\sigma_{p} \sim$ 0.1\% (See Appendix~A1 of \citep{leonard_is_2001}).

Before day +2.5, both the Balmer and the \ion{He}{2} $\lambda$4686 lines are characterized by a narrow emission core in their central region. For a more detailed discussion of the H$\alpha$ core in SN\,2023ixf, refer to the presentation of high-resolution spectra by \citet{2023arXiv230607964S}. This core is superimposed on an electron-scattering wing which can be modeled with a Lorentzian profile. The recombination front of \ion{He}{2} is expected to be deeper than that of H$\alpha$, as the former requires a higher electron temperature (see, e.g., \citet{groh_early-time_2014}). If the recombination front is situated above the photosphere, the photons radiating from the central spectral profile will undergo minimal scattering. As a result, the scattering wings will progressively depolarize toward the rest wavelength of the emission line until they reach a zero level of polarization. For instance, if the recombination front of H$\alpha$ is above the photosphere, complete depolarization will be measured at the center of the H$\alpha$ core assuming that there is no overlap from nearby lines (e.g., \ion{He}{2}). The \ion{He}{2} front could be close to or within the electron-scattering photosphere, thereby producing a more polarized flux due to intensified electron scattering.

The depolarization behavior toward the center of the emission lines is apparent across both \ion{He}{2} $\lambda$4686 and H$\alpha$ at days +1.4 and +2.5. Nevertheless, owing to a minimum resolution of 18\,\AA\ with the chosen Kast spectropolarimeter configuration (equivalent to $\sim 1000$\,km\,s$^{-1}$ in velocity), the measured line profiles are not sufficient to resolve any complete depolarization at the emission-line cores.  Therefore, our low-resolution spectropolarimetry limits our determination of the relative locations between the photosphere and the recombination fronts of H$\alpha$ and \ion{He}{2}.  In any case, the scattering wings of the flash features arise from the CSM; thus, the polarimetry measured across the wings traces the geometric distribution of the CSM.

At day +1.4, as illustrated by the polarized flux profiles $|q|\times f_{\lambda}$ and $|u|\times f_{\lambda}$ in Figure~\ref{fig:pol_ep1}, we detected a slight deviation in the polarized flux across \ion{He}{2} $\lambda$4686 + \ion{N}{3} $\lambda\lambda$4634, 4641 compared to that taken from the adjacent continuum. However, no such deviation can be observed across H$\alpha$. The higher deviation across the former flash feature can be attributed to the recombination fronts of \ion{He}{2} and \ion{N}{3} being located deeper within the CSM compared to H$\alpha$. This is also consistent with the rapid disappearance of \ion{N}{3} on day +1.4, possibly due to an increasing temperature and therefore ionization level in the emitting region of the CSM or to being engulfed by the SN ejecta. 
%\textcolor{purple}{SV2YY: Isn't this also the result of a dropping temperature rather than material being swept out? YY: You are right. Now sure why I wrote that before. N should also extended to the outermost layer of the CSM as it mixed to the progenitor surface before its core-collapses. Only the lowest ionization states such as Ha a better tracer of the emitting region of the ejecta. Please check the above text now} \textcolor{purple}{SV2YY: Thanks Yi! I think the text looks good in the paragraph above now.}. 
However, we caution that the detection of such deviation is marginal. It is also possible that both \ion{He}{2} $\lambda$4686 + \ion{N}{3} $\lambda\lambda$4634, 4641 and H$\alpha$ may be polarized the same as the continuum.

%At day +1.4, as illustrated by the polarized flux profiles ($|q| \times f_{\lambda}$ and $|u|\times f_{\lambda}$ in Figure~\ref{fig:pol_ep1} YY2SV: Please update the figures), we marginally detected a deviation of the polarized flux across \ion{He}{2}$\lambda$4686+\ion{N}{3}$\lambda$4634, 4641 compared to that measured from the adjacent continuum. On the contrary, no such deviation can be seen across H$\alpha$. 
%The higher deviation across the former flash feature can be understood as the recombination fronts of \ion{He}{2} and \ion{N}{3} are located deeper inside the CSM compared to that of H$\alpha$. 
%This is also compatible with the rapid disappearance of the \ion{N}{3} at day +1.4, since the \ion{N}{3}-rich layer, which resides in the innermost regions of the CSM, was first swept by the expanding ejecta. 
%However, we remark that the detection of such deviation is marginal. It is also possible that both \ion{He}{2}$\lambda$4686+\ion{N}{3}$\lambda$4634, 4641 and H$\alpha$ may polarized the same as the continuum.

%On day +2.5, both the scattering wings of the \ion{He}{2}$\lambda$4686 and H$\alpha$ exhibit an increased deviation in $|q|\times f_{\lambda}$ and $|u|\times f_{\lambda}$ from the continuum compared to day +1.4 (see, Figures~\ref{fig:pol_ep1} and \ref{fig:pol_ep2}). 
%Simultaneous to the significant decreasing of the central emission peak, the $|q|\times f_{\lambda}$ and $|u|\times f_{\lambda}$ profiles of \ion{He}{2}$\lambda$4686 and H$\alpha$ have both broadened to FWHM$\sim$2,000 km s$^{-1}$ on day +2.5. 
On day +2.5, the scattering wings of both \ion{He}{2} $\lambda$4686 and H$\alpha$ show an increased deviation in $|q|\times f_{\lambda}$ and $|u|\times f_{\lambda}$ from the continuum, compared to what was observed on day +1.4 (see Figures~\ref{fig:pol_ep1} and \ref{fig:pol_ep2}). This is concurrent with a significant decrease in the central emission peak. Moreover, the $|q|\times f_{\lambda}$ and $|u|\times f_{\lambda}$ profiles of \ion{He}{2} $\lambda$4686 and H$\alpha$ have both broadened to a FWHM of $\sim 2000$\,km\,s$^{-1}$ on day +2.5.
The increased deviation of polarized flux from the continuum becomes even more prominent on day +3.5 (see Figure~\ref{fig:pol_ep3}. 
Such an excess of polarized photons, which exhibits a rather broad profile compared to the electron-scattering wings seen on day +1.4, %might be attributed to the SN ejecta gradually emerging from the CSM. 
%The contribution of electron-scattered photons increases as most of the CSM has been swept away by the expanding ejecta, thus entering a more polarization-efficient, optically thin regime. 
%\textcolor{red}{YY: Perhaps replacing `might be attributed to...regime with the following: 
might be attributed to the recession of the recombination fronts relative to that of the photosphere. As time proceeds, the recession of the recombination fronts for \ion{He}{2} and H$\alpha$ will lead to increased scattering optical depths for both lines. The former is always located at a deeper layer compared to the latter owing to a higher required electron temperature. Consequently, the scattering of the recombination photons from both \ion{He}{2} and H$\alpha$ lines becomes stronger as they emerge from the photosphere. This may account for the progressively enhanced amount of polarized photons measured across the scattering wings observed before $\sim +4$ days, which is manifested as significant departures from that measured at the continuum. This mechanism could take place before the CSM has been overtaken by the expanding ejecta.
We remark that the ejecta expanding into the CSM does not contradict the recession of the recombination front. The latter expands more slowly compared to the former, so both are expanding relative to the CSM, which moves at $\sim 100$\,km\,s$^{-1}$.
It is noteworthy that the H$\beta$ line does not show the $PA$ rotation observed in H$\alpha$ and He~II lines on day 3.5. This is perhaps related to the details of the formation of the hydrogen Balmer lines. The Balmer decrement is measured to be $\sim 1.5$ on day 3.5, which is significantly different from typical values of 2.8 for gas at temperatures of $\sim 10,000$\,K \citep{Miller:1974ARA&A..12..331M}. This suggests that the Balmer lines are formed at a density that is typically optically thick to the Balmer lines such that collisional excitation, radiative excitation by the photospheric continuum, or Balmer-line self-absorption are important. The region of H$\beta$ line formation is expected to more closely track the location of the photosphere than the H$\alpha$ line, thus  maintaining a polarization {\it PA} that more closely follows that of the continuum at day 3.5.

The transition phase between the earlier ``flash" phase and the later ``photospheric" phase occurs over a 2\,day timescale. A significant alteration in geometry was recorded as the CSM was engulfed by the expanding ejecta. This change has been inferred from the temporal evolution of the polarization from days +2.5 to +4.6 (Figure~\ref{fig:story}), which then stabilized to $\sim 0.5$\% along the $u$ direction at later epochs (for instance, as measured at days +9.5 and +14.5).
Polarimetry obtained at day +3.5 samples the geometry during the transitional phase, when the ejecta just surpass the outer boundary of the dense CSM. Meanwhile, polarimetry at day +4.6 samples the geometry of the SN ejecta after they have fully overtaken the dense CSM, as indicated by the disappearance of the flash features.

On day +4.6, no significant deviation can be identified in the polarized flux across both \ion{He}{2} $\lambda$4686 and H$\alpha$ (see Figures~\ref{fig:pol_ep4}. 
The former disappeared after day +9.5, and the latter evolved into a weak and broader P~Cygni profile, suggesting that the SN ejecta have completely swept up the CSM. 
As the photosphere recedes inside the ejecta, the constant $p \approx 0.5$\% and $PA \approx 220^{\circ}$ after day +4.6 thus reflect the geometry of the SN ejecta.
The constant $PA$ obtained from days +4.6 to +14.5 is also consistent with the 
continuum polarization settling to 0.5\%, while also suggesting that the outermost H-rich layer of SN ejecta is itself aspherical.
%\textcolor{blue}{RC: Here might be the place to revisit the question of ISP. ISP cannot explain the features you describe in this section, but removing a small amount of ISP consistent with the other constraints would change the exact values of the PA, etc. }

 Although we do not address the removal of ISP, our general interpretations remain unchanged, even at the upper limit of $p_{\rm ISP,max} = 0.35$\%. This is because ISP cannot account for the rapid decrease in continuum polarization and the $PA$ rotation. A proper determination and removal of the ISP would result in a shift of the origin in the $q-u$ plane (see blue dashed circle in Figure \ref{fig:story}). 
%This is where I can reference that 0.35\% circle.
However, the exact values of the continuum polarization and $PA$ will differ depending on $q_{\text{ISP}}$ and $u_{\text{ISP}}$.

We also remark that the scenario of the early ejecta-CSM interaction depicted by the temporal evolution of the polarimetric behavior is generally consistent with the physical picture derived from the evolution of spectral-profile widths. These profiles arise from different regions of the CSM and the interface where the SN ejecta interact with the CSM, as discussed by \citet{2023arXiv230607964S}. 
The highly aspherical mass-loss process leading up to the explosion could have been shaped by interactions between eruptive mass ejections and a binary companion, as suggested by pre-SN modeling \citep{soker_binary_2013,2014ApJ...785...82S}. %\textcolor{purple}{SV: I need to change the wording here. Something sounds off. Maybe: "The intense mass-loss process leading up to the explosion is characterized by a highly aspherical configuration. Such a process is likely triggered by interactions between eruptive mass ejections and a binary companion, as suggested by pre-SN hydrodynamical modeling \citep{2014ApJ...785...82S}."}\textcolor{blue}{YY: Sounds OK for now.} SV: FIXED!
The detailed geometric properties of the SN ejecta, which may have been shaped by turbulent processes during the progenitor's final burning stages leading to the explosion or by asymmetries in the explosion process itself, still remain unclear. 
%Therefore, the different geometries as inferred from the earlier `flash' and the later `photospheric' phase may not necessarily suggest that the CSM enrichment and the SN explosion are two unrelated procedures. 
The geometry of the core and information about the preferential axis of the explosion could be further probed by monitoring the temporal evolution of the continuum polarization after SN\,2023ixf falls off from its plateau phase. Also, inspection of the emission-line profiles during the nebular phase will provide an additional probe of any asymmetries in the CSM.
%Observationally, the role played by a binary companion during the explosion phase could be further probed by monitoring the temporal evolution of the continuum polarization after SN\,2023ixf falls off from its plateau phase, or a careful inspection of the spectral profiles at such late phases, when the recombination front has receded inside the H-rich envelope so the inner ejecta becomes transparent.
%\textcolor{purple}{SV2YY: I am not entirely sure If I understand the logic behind the last sentence [*paragraph]. Perhaps we can discuss this tomorrow?}
%\textcolor{red}{YY2SV: Not sure what else I can do.}
%\textcolor{purple}{SV2YY: I tried to fix the paragraph. Will probably need feedback from others...}

%\textcolor{purple}{SV2YY: I still want to add discussion on the geometries possible.  For example, something along the lines of: Disk-like, toroidal geometry with fall off in density past 10$^{15}$ cm. The spectropolarimetry observations along with the modeling of the flash ionization emission spectra [not very blueshifted absoprtion components of p-cygni recombination profilessmith+23 paper] most likely favor a disk-like or toroidal distribution of the CSM. }
%\textcolor{red}{YY2SV: I'm not sure if I have anything to say about this, please go ahead if you want to discuss more.}
The physical scenario presented in Figure \ref{fig:schem} is also compatible with SN ejecta expanding into a disk-like, toroidal geometry characterized by a steep falloff in density beyond a CSM radius $R_{\text{CSM}} \approx 10^{15}$\,cm. The SN ejecta break out through the higher latitudes and are pinched in the plane of the disk. After +4.6 days, the SN ejecta engulf the dense CSM disk/torus and are elongated in the polar directions. A similar model was used to describe the rapid rotation of the $PA$ and the change in continuum polarization for the interacting Type II SNe 1997eg \citep{hoffman_dual-axis_2008} and 2009ip (see Figures 9 and 10 of \citet{mauerhan_multi-epoch_2014}), albeit at much later times.  The spectropolarimetric observations presented in this work align with the conclusions of  \citet{2023arXiv230607964S}, where the CSM was also suggested as being asymmetric from the lack of blueshifted absorption components in the narrow P~Cygni profiles. Another possible model for the CSM includes the presence of a pre-explosion effervescent zone (see \citet{soker_pre-explosion_2023}). The polarimetry results suggest that such an effervescent zone may be flat due to a rotation of the progenitor or binary interaction.

%\textcolor{blue}{RC: This section needs some comparison to the SN IIn specpol literature. In addition to the Leonard et al. paper on 98S commented on earlier, there is the recent compilation by Chris Bilinski: https://ui.adsabs.harvard.edu/abs/2023arXiv230413034B/abstract and older works on objects such as 10jl https://ui.adsabs.harvard.edu/abs/2011A\%26A...527L...6P/abstract and possibly 97eg: https://ui.adsabs.harvard.edu/abs/2008ApJ...688.1186H/abstract These papers demonstrate that depolarizations across the line features are typical of interacting SNe, just like what you see. Do you want to make any comments about the structure of the CSM environment to contrast ``flash" SNe II with the more extended CSM seen in SNe IIn? }
%\textcolor{purple}{SV2RC: Yes, thank you for suggesting this. I agree, it is very important that we at least mention 98S in this paragraph. I will write this up. I added SN 1997eg to previous paragraph} \\ \\

We draw comparisons between our polarimetric observations and those of other interacting SNe such as Type IIn SN\,1998S \citep{shivvers_early_2015}. In contrast to SN\,2023ixf, SN\,1998S exhibited prolonged ``flash" features ($> 5$\,days) and a high ($p \sim 3$\%) continuum polarization \citep{leonard_evidence_2000} at early times, a characteristic commonly observed in SNe\,IIn, indicative of the presence of an extended CSM \citep{bilinski_multi-epoch_2023}. The depolarization observed near the emission-line cores of SN\,2023ixf have been observed in SNe\,IIn such as SN\,1997eg \citep{hoffman_dual-axis_2008} and 2010jl \citep{patat_asymmetries_2011}, which were also characterized by a high continuum polarization similar to that of SN\,1998S. Models of these interacting SNe favor an extended, dense CSM with distinct narrow-line regions in an equatorial plane (disk) and a diffuse broad-line region which dominates in the polar directions  \citep{leonard_evidence_2000,hoffman_dual-axis_2008}.  Other models, compatible with the equatorial CSM distribution, favor a cold dense shell (CDS) envelope producing narrow-line emission, surrounded by a broad-line region \citep{chugai_type_2004,dessart_models_2016}. 
In the observed sample of interacting SNe, a pattern emerges: persistent asymmetries exist with a certain diversity that can be explained by viewing-angle effects \citep{bilinski_multi-epoch_2023}, as well as varying environments on a continuum, which likely involve either binary interaction or eruptive mass-loss processes.
%The temporal evolution of the polarimetric behavior depicts a scenario that in general agrees with that inferred from the evolution of various widths of spectral profiles that arises from various regions of the CSM and the ejecta-CSM interacting zones as discussed in \citet{2023arXiv230607964S}. 
%Such an intense mass-loss procedure that follows a highly aspherical configuration is most likely caused by rather immediate interactions between eruptive mass ejections and a binary companion \citep{2014ApJ...785...82S}. 
%The detailed geometric properties of the SN ejecta that may be sculpted by the effect of turbulence during the final stages of the progenitor burning that may lead to the terminated explosion still remain unclear. 
%Therefore, the different geometries as inferred from the earlier `flash' and the later `photospheric' phases may not necessarily suggest that the CSM enrichment and the SN explosion are two unrelated procedures.  
%Observationally, the role played by a binary companion during the explosion phase could be further probed by monitoring the temporal evolution of the continuum polarization after SN\,2023ixf falls off from its plateau phase, or a careful inspection of the spectral profiles at such late phases, when the recombination front has receded inside the H-rich envelope so the inner ejecta becomes transparent.

%\hspace{}
\section{Conclusions}\label{sec:conclusion}
We present a spectropolarimetric sequence of the nearby Type II SN\,2023ixf carried out starting 1.4\,days after the SN explosion. The first epochs of observations (days +1.4 and +2.5), which were obtained before the disappearance of a series of highly ionized emission features, are the earliest such measurements for any SN to date.
The timing of these observations, combined with their high cadence, offers an unprecedented opportunity to investigate the ``flash" ionization features, the outermost layers of the explosion, and its immediate circumstellar environment. 
%This high-cadence tomography, initiated shortly after the SN explosion, offers an unprecedented opportunity to determine the geometric configuration during the earliest ejecta-CSM interaction phase. 
The inferred geometric properties can therefore further constrain % the immediate circumstellar environment and 
the associated progenitor mass-loss history.

A high continuum polarization ($p\approx$1\%) was measured at days +1.4 and +2.5 after the explosion, compatible with a highly aspherical, optically thick CSM that extends beyond the location of the shock-driven SN ejecta at this phase.
A constant $PA \approx 165^{\circ}$ observed across both the continuum and the scattering wings of the short-lived, narrow emission lines (FWHM $\approx 1000$\,km\,s$^{-1}$) may indicate a common emitting region shared by both the continuum and the early recombination features, namely the outer layers of the ambient CSM.
On day +3.5, the continuum polarization dropped to nearly 0.5\%. The polarization angle of the continuum remained near $\sim 165^{\circ}$ to within the systematic uncertainties. Meanwhile, the $PA$ along the wings of the H$\alpha$ and \ion{He}{2} $\lambda$4686 lines rotated gradually to $\sim 225^{\circ}$ and $\sim 200^{\circ}$ (respectively) toward their rest wavelengths. This rapid change is observed as a shift of the continuum polarization in the Stokes $q-u$ plane, coinciding with the weakening of the ``flash" features.   
%\textcolor{red}{YY: `
%At day +3.5, when the flash features are vanishing quickly, a conspicuous shift of the continuum polarization has occurred as viewed on the Stokes $q-u$ plane, which is concurrent with a significant change of the polarization profile across the broadened H$\alpha$ line and the \ion{He}{2}$\lambda$4686 feature. The $PA$ was measured to be $\approx165^{\circ}$ over the continuum but has evolved into a gradually increased profile towards $\sim$225$^{\circ}$ as measured close to the inner wings of these emission lines.'}
After day +4.6, the continuum polarization has settled to a relatively constant value in the Stokes $q-u$ plane, with the corresponding $p$ and $PA$ measured to be approximately 0.5\% and 225$^{\circ}$, respectively. The $PA$ again shows a constant level across both the continuum and the major emission lines.

Based on the temporal evolution of the polarization of SN\,2023ixf, we propose a physical scenario where the aspherical, optically thick CSM is swept away by the aspherical SN ejecta within the first five days after the explosion of the progenitor star as illustrated in Figure~\ref{fig:schem}.
The asymmetry in the CSM likely results from a disk-like or toroidal geometry produced by eruptive mass loss occurring within a few years before the explosion, possibly involving interaction with a binary companion. Ongoing spectropolarimetric surveillance of SN\,2023ixf through the falloff from its photospheric plateau and during nebular phases will provide valuable insights into the origins of the observed asymmetries.
%\begin{appendix}
%\end{appendix}

\section{acknowledgments} 
A major upgrade of the Kast spectrograph on the Shane 3\,m telescope at Lick Observatory, led by Brad Holden, was made possible through gifts from the Heising-Simons Foundation, William and Marina Kast, and the University of California Observatories.
% KAIT and its ongoing operation were made possible by donations from Sun Microsystems, Inc., the Hewlett-Packard Company, AutoScope Corporation, Lick Observatory, the U.S. National Science Foundation, the University of California, the Sylvia \& Jim Katzman Foundation, and the TABASGO Foundation. 
We appreciate the expert assistance of the staff at Lick Observatory.  
Research at Lick Observatory is partially supported by a gift from Google. 

Generous financial support was provided to A.V.F.'s supernova group at U.C. Berkeley by the Christopher R. Redlich Fund, Steven Nelson, Landon Noll, Sunil Nagaraj, Sandy Otellini, Gary and Cynthia Bengier, Clark and Sharon Winslow, Sanford Robertson, Alan Eustace, Briggs and Kathleen Wood, and numerous other donors.
R.M.\ acknowledges support by the National Science Foundation (NSF) under awards AST-2221789 and AST-2224255. L.W. acknowledges the NSF for support through award AST-1817099. The TReX team at U.C. Berkeley is partially funded by the Heising-Simons Foundation under grant \#2021-3248 (PI R. Margutti).
\bigskip

%This research made use of \texttt{TARDIS}, a community-developed software package for spectral synthesis of SNe \citep{kerzendorf_spectral_2014, kerzendorf_wolfgang_2022_6299948}. The development of \texttt{TARDIS} received support from GitHub, the Google Summer of Code initiative, and ESA's Summer of Code in Space program. \texttt{TARDIS} is a fiscally sponsored project of NumFOCUS. \texttt{TARDIS} makes extensive use of Astropy and Pyne.

\software{{Astropy \citep{astropy:2013, astropy:2018}, 
%\texttt{TARDIS} \citep{kerzendorf_spectral_2014,vogl_spectral_2019}, 
%uvotpy \citep{kuin_uvotpy_2014}, 
%DAOPHOT \citep{stetson_daophot_1987}, 
IDL Astronomy user's library \citep{landsman_idl_1993}}} 
%SOUSA pipeline \citep{brown_sousa_2014}, }
\bigskip
\newpage
\appendix
\section{Appendix: Polarization Spectra of SN\,2023ixf~\label{sec:app}}
\restartappendixnumbering

%YY20230701\begin{figure}[h]
%    \centering
%    \includegraphics[width=0.5\linewidth]{figures/2023ixf_ep2_final_quflux20_v2.pdf}
%     \caption{Same as Figure~\ref{fig:pol_ep1}, but for day $+$2.5 (epoch 2).}~\label{fig:pol_ep2}
%\end{figure}

%\begin{figure*}
%   \begin{minipage}[t]{0.50\textwidth}
%     \centering
%     \includegraphics[width=1.0\linewidth]{figures/2023ixf_akj.pdf}
%     \caption{
%     Spectropolarimetry of SN\,2023ixf on day +1.4 (epoch 1) relative to an estimated time of the first light on MJD 60082.83. The five panels (from top to bottom) present (a) the arbitrarily scaled flux density with major spectral features labeled; (b,c) the fractional Stokes parameters $q$ and $u$ , respectively; (d) the degree of polarization ($p$); and (e) the polarization position angle ($PA$). In panels (b)–(e), the data have been rebinned to 30\AA\ for clarity. No ISP correction has been applied. Grey-shaded vertical bands mark regions of telluric contamination. 
%     }~\label{fig:pol_ep1}
%   \end{minipage}\hfill
%   \begin{minipage}[t]{0.50\textwidth}
%     \centering
%     \includegraphics[width=1.0\linewidth]{figures/2023ixf_ep220.pdf}
%     \caption{Same as Figure~\ref{fig:pol_ep1}, but for day $+$3 (epoch 2).}~\label{fig:pol_ep2}
%   \end{minipage}
%\end{figure*}

\begin{table*}[!h]
\begin{center}
\caption{Journal of spectropolarimetric observations of SN\,2023ixf.} 
    \begin{tabular}{cccccccc}
	\hline 
	\hline
	UT Date & MJD$^b$ & Phase$^a$ & Airmass & Avg. Seeing &  Exposure Time$^c$   \\ 
	(MM-DD-YYYY)&   & (days) & & (arcsec) & (s)   \\ 
	\hline 
%    09-11-2021 && & Kast & 1.9-2.1 & 3 & 3 & 360 \\
    05-20-2023 & 60084.18 & 1.4 & 1.12--1.44 & 1.2 & $600 \times 4 \times 4$ \\
    05-21-2023 &60085.23& 2.5 & 1.06& 1.2 & $600 \times 4 \times 3$ \\
    05-22-2023 &60086.24& 3.5 & 1.05 & 1.2 &  $600 \times 4 \times 1$\\
    05-23-2023 &60087.35& 4.6 & 1.15 & 1.2 &  $600 \times 4 \times 1$ \\
    05-28-2023 &60092.20& 9.5 & 1.05 & 1.2 &  $600 \times 4 \times 4$ \\
   06-02-2023 &60097.26& 14.5 & 1.06 & 1.2 &  $200 \times 4 \times 2$\\
	\hline 
\end{tabular}\\
%\flushleft
{$^a$}{Days after the estimated time of first light on MJD~60082.75 (UTC 18 May 2023).} \\
{$^b$}{MJD is given as the start time of the CCD exposure.}\\
{$^c$}{Exposure time of a single exposure $\times$ 4 retarder-plate angles $\times$ number of loops. Wavelength range for Kast is 4550--9800\,\AA.} 
\label{tbl:specpol_log}
\end{center}
\end{table*} 

\begin{figure*}[!h]
   \begin{minipage}[t]{0.50\textwidth}
     \centering
     \includegraphics[trim={0.5cm 2.5cm 0.5cm 2.5cm},clip,width=1.0\textwidth]{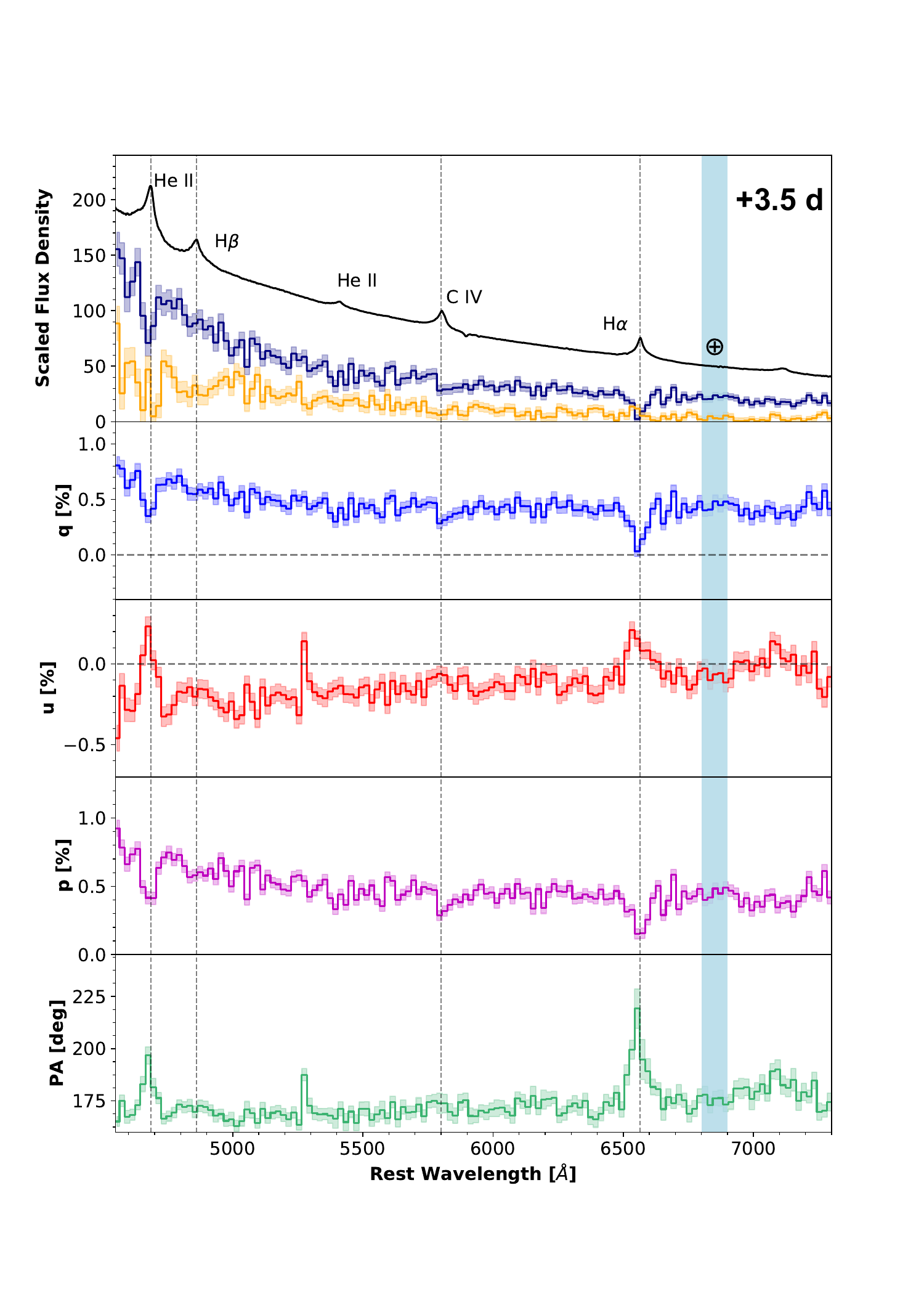}
    \caption{Same as Figure~\ref{fig:pol_ep1}, but for day $+$3.5 (epoch 3).
    }~\label{fig:pol_ep3}
   \end{minipage}\hfill
   \begin{minipage}[t]{0.50\textwidth}
     \centering
     \includegraphics[trim={0.5cm 2.5cm 0.5cm 2.5cm},clip,width=1.0\linewidth]{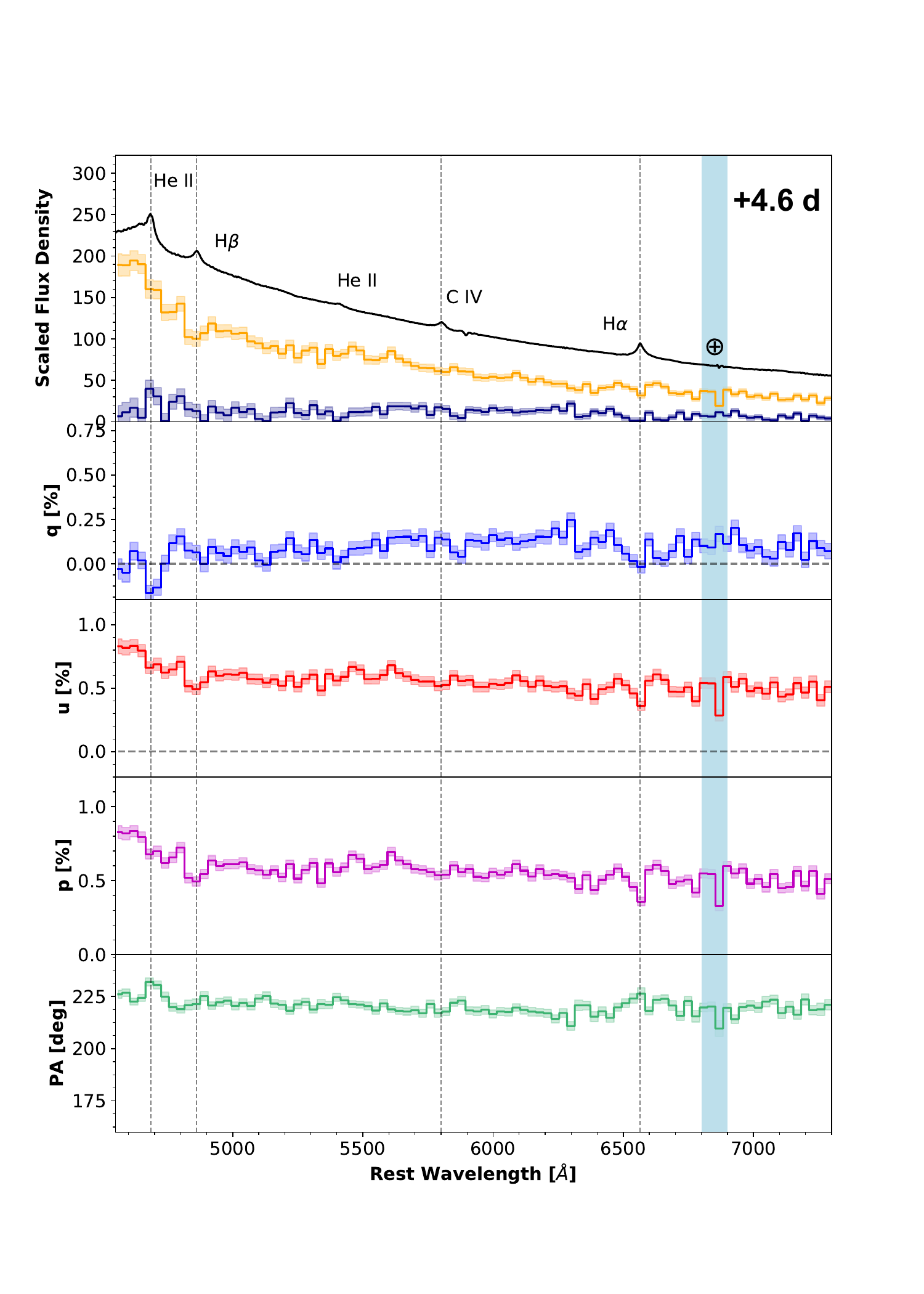}
     \caption{Same as Figure~\ref{fig:pol_ep1}, but for day $+$4.6 (epoch 4).}~\label{fig:pol_ep4}
   \end{minipage}
\end{figure*}
\clearpage

\begin{figure*}[!h]
   \begin{minipage}[t]{0.50\textwidth}
     \centering
     \includegraphics[trim={0.5cm 2.5cm 0.5cm 2.5cm},clip,width=1.0\linewidth]{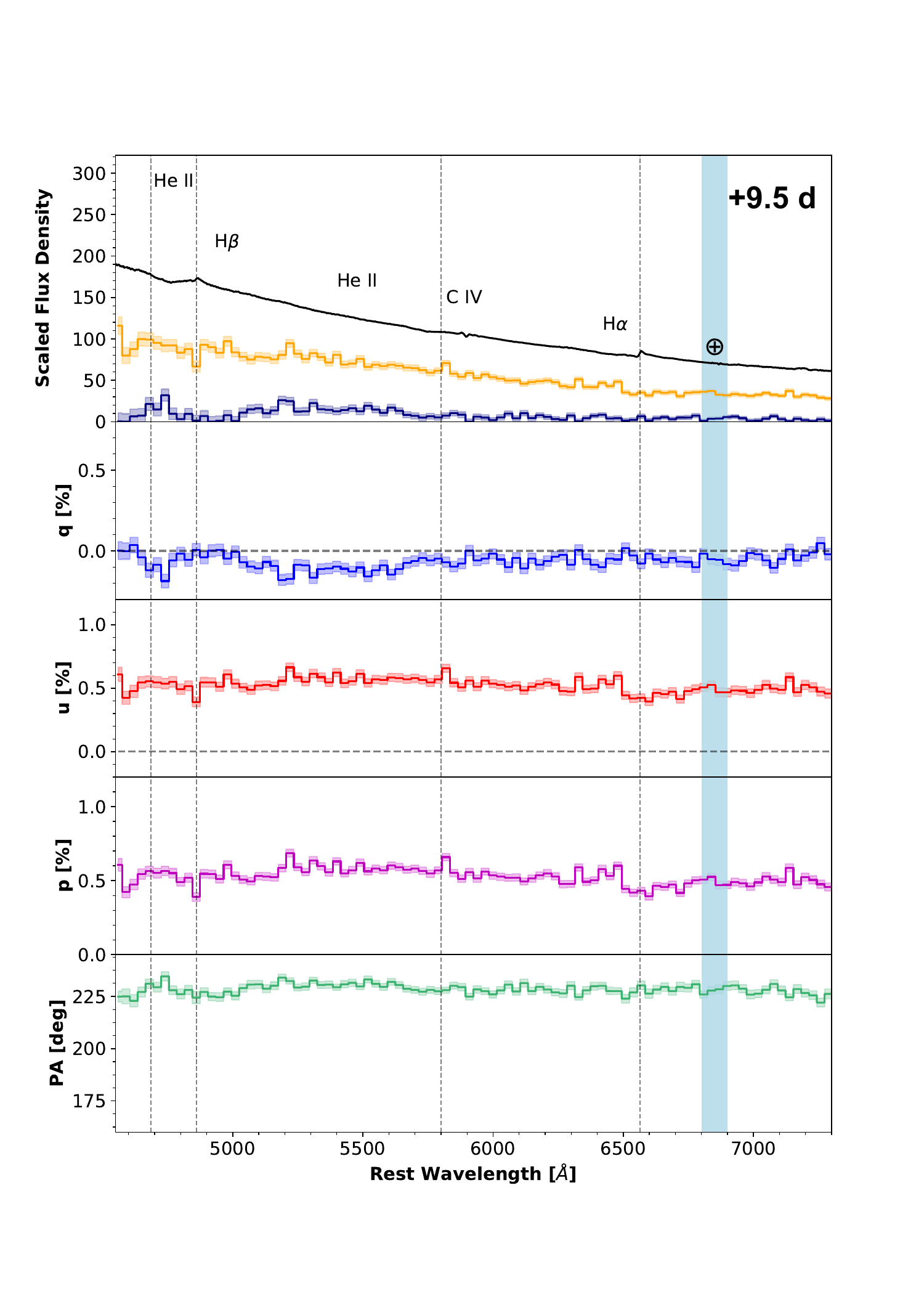}
     \caption{Same as Figure~\ref{fig:pol_ep1}, but for day $+$9.5 (epoch 5).}~\label{fig:pol_ep5}
   \end{minipage}\hfill
   \begin{minipage}[t]{0.50\textwidth}
     \centering
     \includegraphics[trim={0.5cm 2.5cm 0.5cm 2.5cm},clip,width=1.0\linewidth]{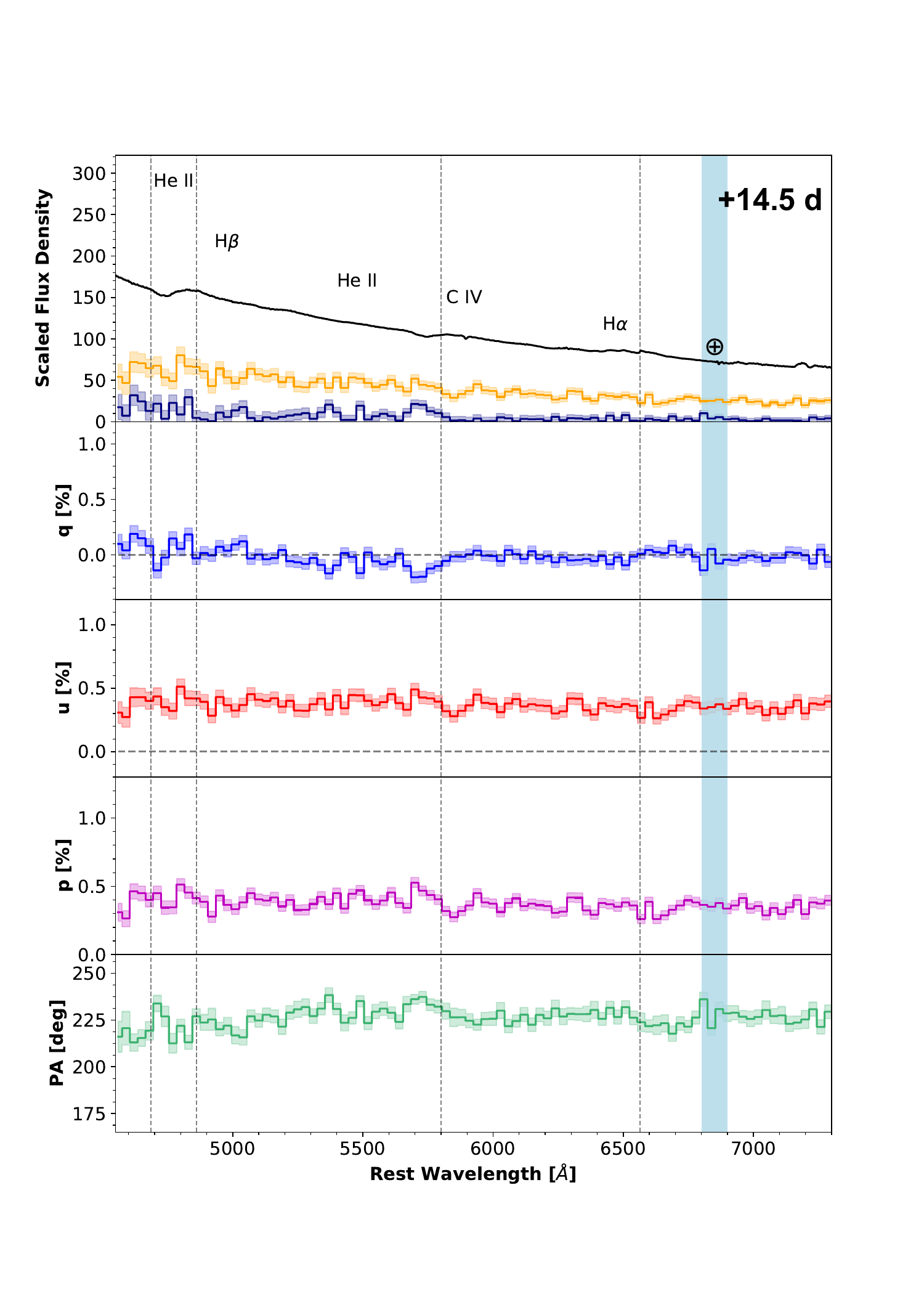}
     \caption{Same as Figure~\ref{fig:pol_ep1}, but for day $+$14.5 (epoch 6).}~\label{fig:pol_ep6}
   \end{minipage}
\end{figure*}
\begin{figure*}
    \centering
    \includegraphics[width=1.0\textwidth]{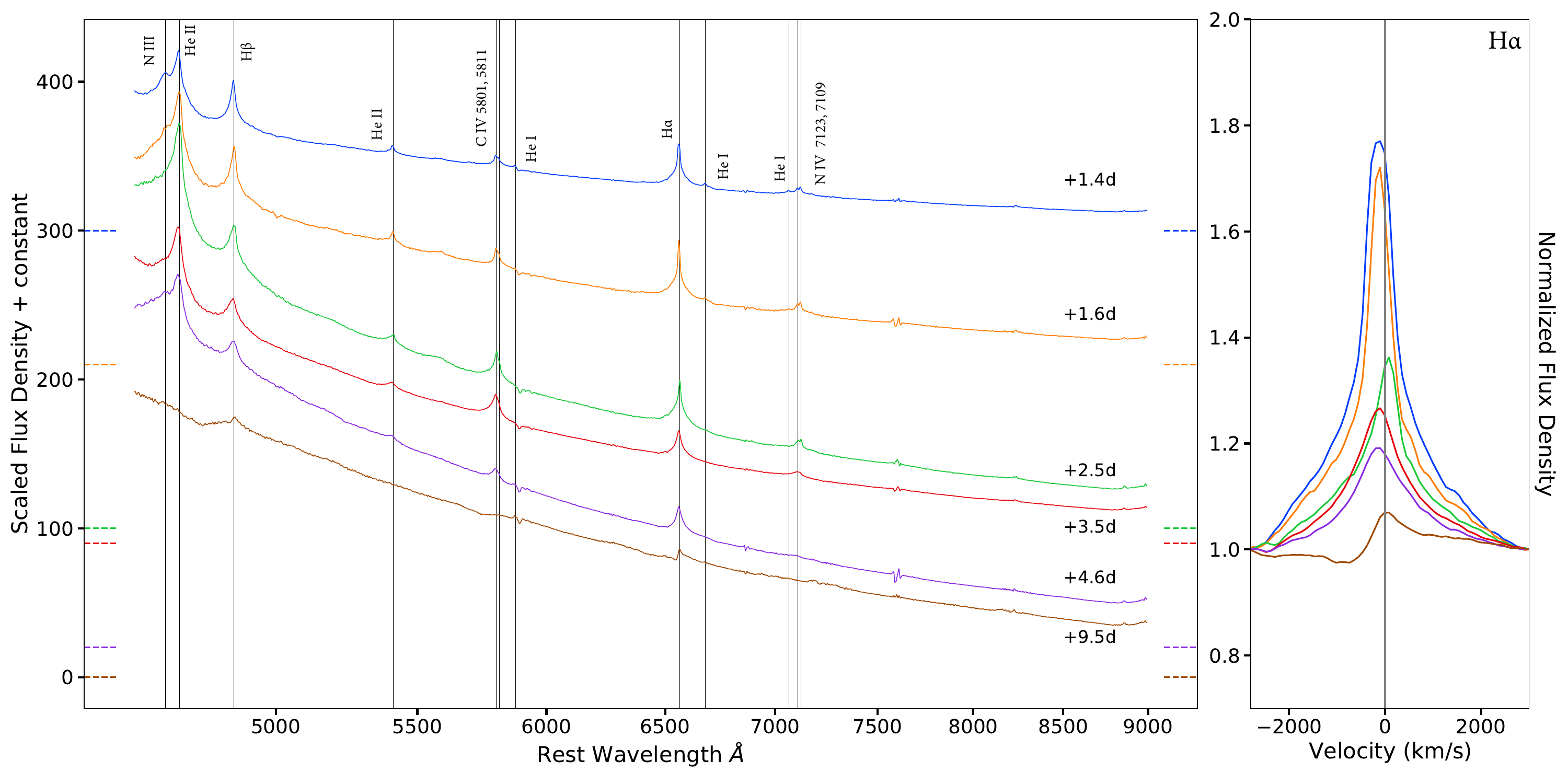}
    \caption{{\it Left:} Temporal evolution of the flux spectra (Stokes $I$) of SN\,2023ixf presented in this paper. A comparison between the two sets of observations obtained at the first epoch, namely MJD 60084.20 (day +1.4) and 60084.39 (day +1.6) illustrates the evolution of \ion{N}{3} lines on a $\sim$ 1 hour timescale. Vertical black solid lines indicate prominent emission features. Dashed horizontal lines indicate the zero point of each flux spectrum with the corresponding color. Each spectrum is offset by a constant with the exception of day +9.5. The flux spectra on days +1.4, +2.5, and +3.5 are scaled by a factor of four, four, and  two, respectively. Days +4.6 and +9.5 are presented without any scaling to the flux.
    {\it Right:} A zoomed-in view of the H$\alpha$ profile of SN\,2023ixf. The flux density is normalized relative to the continuum.}
    \label{fig:fluxes}
\end{figure*}
%\clearpage

\bibliographystyle{aasjournal}
\bibliography{2023ixf}

\listofchanges
\end{document}